\documentclass[twocolumn,showpacs,amsmath,amssymb,prd,floatfix,nofootinbib,preprintnumbers,longbibliography]{revtex4}
\usepackage[dvipdfmx]{graphicx}
\usepackage{bm}
\usepackage{amsmath}
\usepackage{color}
\usepackage{xcolor}
\usepackage{hyperref}
\usepackage{physics}
\usepackage{comment}
\usepackage{multirow}
\usepackage{url}
\newcommand{\bk}{{\bf k}}

\newcommand{\tW}{\tilde{W}}
\newcommand{\tu}{\tilde{u}}

\newcommand{\del}{\partial}

\newcommand{\KCDM}{\mbox{$\Omega_K$-$\Lambda$CDM}}

\begin{document}

\title[]{Separate universe approach to evaluate nonlinear matter power spectrum \\
for non-flat $\Lambda$CDM model}

\author{Ryo~Terasawa$^{1,2}$}
\email{ryo.terasawa@ipmu.jp}
\author{Ryuichi~Takahashi$^3$}
\author{Takahiro~Nishimichi$^{4,1}$}
\author{Masahiro~Takada$^1$}
\affiliation{%
$^1$Kavli Institute for the Physics and Mathematics of the Universe (WPI),
 The University of Tokyo Institutes for Advanced Study (UTIAS),
 The University of Tokyo, Chiba 277-8583, Japan
}%
\affiliation{%
$^2$Department of Physics, Graduate School of Science,
 The University of Tokyo, 7-3-1 Hongo, Bunkyo-ku, Tokyo 113-0033, Japan
}%
\affiliation{
$^3$Faculty of Science and Technology, Hirosaki University, 3 Bunky
o-cho, Hirosaki, Aomori 036-8561, Japan
}
\affiliation{
$^4$Center for Gravitational Physics and Quantum Information, Yukawa Institute for Theoretical Physics, Kyoto University, Kyoto 606-8502, Japan
}

\begin{abstract}
The spatial curvature ($\Omega_K$) of the Universe is one of the most fundamental quantities that 
could give a link to the early universe physics. In this paper we develop an approximate method to compute the nonlinear matter power spectrum, $P(k)$, for ``non-flat'' $\Lambda$CDM models using the separate universe (SU) 
ansatz which states
that the effect of the curvature on structure formation is equivalent to that of long-wavelength density fluctuation ($\delta_{\rm b}$)
in a local volume in the ``flat'' 
$\Lambda$CDM model, via the specific mapping between the background cosmological parameters and redshifts in the non-flat and flat models. 
By utilizing the fact that the normalized response 
of $P(k)$ to $\delta_{\rm b}$ (equivalently $\Omega_K$), which describes how the non-zero $\Omega_K$ alters $P(k)$ as a function of $k$, 
is well approximated by the response to the Hubble parameter $h$ within the flat model,
our method allows one to generalize the prediction of $P(k)$ for flat cosmologies via fitting formulae or emulators to that for 
non-flat cosmologies.
We use $N$-body simulations for the non-flat $\Lambda$CDM models with $|\Omega_K|\leq 0.1$
to show that our method can predict $P(k)$ for non-flat models up to $k \simeq 6\,h{\rm Mpc}^{-1}$ in the redshift range $z\simeq [0,1.5]$, to the fractional accuracy within $\sim 1$\% 
that roughly corresponds to requirements for weak lensing cosmology with upcoming surveys. 
We find that the emulators, those built for flat cosmologies such as {\tt EuclidEmulator}, can predict the non-flat $P(k)$ with least degradation.  
\end{abstract}

\preprint{IPMU22-0027,YITP-22-49}
\maketitle

\section{Introduction}

The spatial curvature of the Universe (hereafter denoted as its contribution relative to the present-day critical density, $\Omega_K$) 
is one of the most fundamental quantities in an isotropic and homogeneous universe in the context of General Relativity \cite{1972gcpa.book.....W}.
The curvature also has a close connection to the physics of the early universe. 
An inflationary universe scenario predicts that the {\it apparent} curvature, inferred from an observable universe, should be close to 
a flat geometry ($\Omega_K\approx 0$), even if its exact value is non-zero \cite{1981MNRAS.195..467S,1981PhRvD..23..347G}. 
If the universe arose form the decay of a false vacuum via quantum tunneling, as inspired by the landscape of string cosmology vacua, 
it leads to an open-geometry universe ($\Omega_K>0$) 
\citep{1980PhRvD..21.3305C,1982Natur.295..304G}. Depending on the details of the early universe physics, the curvature can be large enough, such as 
$|\Omega_K|\simeq [10^{-4},10^{-2}]$, to be measurable 
by cosmological observations \citep{2012PhRvD..86b3534G,2012JCAP...06..029K}\footnote{If the curvature is as small as $|\Omega_K|\sim 10^{-5}$, 
we cannot distinguish between the global curvature and the primordial curvature ``perturbations''. Hence, the target goal for a hunt of the non-zero curvature is in the range $|\Omega_K|\gtrsim \mbox{a few}
\times 10^{-5}$.}.
Therefore, an observational exploration of the curvature is an important direction to pursue with ongoing and upcoming cosmology 
datasets \citep[e.g.][]{2014PASJ...66R...1T,2015PhRvD..92l3518T}.

The curvature affects cosmological observables in two ways. First is its geometrical effect.
The most promising observables are the baryon acoustic oscillations (BAO) imprinted onto the cosmic microwave 
background (CMB) anisotropies \citep{1996PhRvD..54.1332J} and the distribution of galaxies \citep{2005ApJ...633..560E}. 
Although the BAO scale \citep[more exactly the sound horizon, as proposed by the pioneer works][]{1970ApJ...162..815P,1970Ap&SS...7....3S}
is set by the physics in the early universe, where the curvature's effect is negligible, the angular extent (and redshift difference) 
of the BAO scale, 
inferred from the CMB and galaxy observables, are determined by light propagation, which in turn allows us to infer the curvature parameter from the measured cosmological distances. 
Secondly, the curvature affects the growth of cosmic structure; since the time evolution of density fluctuation field  arises from competing effects between the gravitational 
pulling force and the cosmic expansion, the curvature leaves a characteristic signature in the growth history of large scale structure via its effect on the cosmic 
expansion \citep[e.g.][]{1999Sci...284.1481B}. 

The geometrical constraint, inferred from 
the primary CMB anisotropy information of the {\it Planck} data \citep{planck_collaboration_2020},
is given as
$\Omega_K=-0.044^{+0.018}_{-0.015}$ (68\% CL, {\it Planck} TT, TE, EE+lowE), implying a $2\sigma$ hint of the close geometry, although the constraint suffers from a severe parameter degeneracy (e.g. with the Hubble parameter $h$).
The joint CMB and galaxy BAO measurements give the tightest constraint on $\Omega_K$, consistent with a flat geometry:
$\Omega_K = -0.0001 \pm 0.0018$ \citep{eBOSS2021}. Ideally we want to use only 
galaxy BAO measurements at {\it multiple} redshifts to constrain the curvature, without employing the CMB prior on the BAO scale, to address whether the CMB and galaxy datasets have consistency within $\Lambda$CDM cosmologies, as motivated by the possible tensions between 
the CMB (early-time) and late-time universe datasets \citep[e.g. see][for the recent review]{2022JHEAp..34...49A}.

On the other hand, the growth constraint on the curvature is still in the early stage. 
Weak lensing and galaxy clustering, observed from 
wide-area galaxy surveys, are powerful methods to constrain cosmological parameters. 
However, most of the previous cosmological analyses assume a flat geometry and focus on the parameters to characterize the clustering
amplitudes such as $S_8$ and $\Omega_{\rm m}$ \citep[e.g. see][for the attempt to constrain $\Omega_K$ from the weak lensing data]{2021A&A...649A..88T}. Although the curvature effect on the linear growth factor 
is accurately known, the linear-regime information is weaker than the BAO constraint. To obtain a tighter constraint on the curvature, we need an accurate model of the clustering observables that are applicable to the nonlinear regime. 
$N$-body and hydrodynamical simulations of cosmic structure formation are among the most powerful, accurate method for such a purpose.
However, simulations are still expensive to construct the theoretical templates, especially in a multi-dimensional 
parameter space such as the vanilla $\Lambda$CDM model plus the curvature parameter.  
A more practical method at this stage is using the fitting formula or ``emulation'' based method \citep[e.g.][]{1996MNRAS.280L..19P,Smith03,Coyote1,Coyote2,Takahashi12,MiraTitan1,MiraTitan2,Nishimichi_2019,2020MNRAS.492.5226R,2021MNRAS.502.1401M,2019MNRAS.486.1448S,2021MNRAS.505.2840E}. 
However, such efforts developing the emulation method are usually done assuming flat-geometry cosmologies due to the computational expense.

Hence the purpose of this paper is to develop, as the first step, an approximate method for computing the nonlinear matter power spectrum, 
$P(k)$, for non-flat cosmologies, which is the fundamental quantity for weak lensing cosmology
\citep{1999ApJ...522L..21H,2004MNRAS.348..897T}.
In fact the existing weak lensing measurements have been used to obtain
tight constraints on the cosmological parameters \citep{2017MNRAS.465.1454H,2018PhRvD..98d3528T,2019PASJ...71...43H,2020PASJ...72...16H,2021A&A...645A.104A,2022PhRvD.105b3515S}.
The current and upcoming weak lensing surveys require a 1\%-level or even better accuracy in the 
theoretical template of $P(k)$ up to $k\sim 1\,h{\rm Mpc}^{-1}$
in order not to have a significant bias in cosmological parameters such as dark energy parameters \citep{Huterer:2004tr}.  
In this paper we employ the separate universe (SU) approach to study $P(k)$ for non-flat $\Lambda$CDM cosmologies. The SU ansatz states that 
the effect of the curvature on structure formation in a given non-flat $\Lambda$CDM model is equivalent to the effect of the long-wavelength (super-box) 
density fluctuation on the evolution of short-wavelength (sub-box) fluctuations in the counterpart flat-geometry $\Lambda$CDM model \cite{2011JCAP...10..031B,2013PhRvD..87l3504T,2014PhRvD..89h3519L,2014PhRvD..90j3530L,2015MNRAS.448L..11W},
where  the cosmological parameters and redshifts in between the non-flat and flat models have to be mapped in the specific way. 
To study structure formation in the two mapped models, it is useful to use the ``response'' function of $P(k)$ which quantifies how $P(k)$ responds to 
the long-wavelength density fluctuation or equivalently the non-zero
curvature, as a function of $k$.
To develop our method, we further utilize the approximate identity that the response of $P(k)$ to 
the curvature, normalized relative to the response in the linear regime, is approximated by the normalized response of $P(k)$ to the Hubble parameter 
$h$ \citep{2014PhRvD..90j3530L}. By using the response to $h$, we can express $P(k)$ for a target non-flat $\Lambda$CDM model in terms of quantities for the corresponding 
flat $\Lambda$CDM model. That is, our method allows us to extend fitting formula or emulator, developed for flat-geometry cosmologies, to predicting $P(k)$ for 
non-flat model, which eases the computational cost for constructing the theoretical templates. 
We will validate our method using a set of $N$-body simulations for flat and non-flat $\Lambda$CDM models with $|\Omega_K|\le 0.1$. We will also assess 
the performance of the publicly available emulator for computing $P(k)$ for non-flat models.

This paper is organized as follows. In Section~\ref{sec:SU_estimator} we first review the SU approach and then 
describe our approximate method for computing the nonlinear matter power spectrum for non-flat $\Lambda$CDM models. 
In Section~\ref{sec:simulations} we describe details of $N$-body simulations for flat and non-flat $\Lambda$CDM models. In Section~\ref{sec:results} 
we present the main results of this paper. We first validate the approximation for the normalized growth response as we described above, and
then show the accuracy of our method for predicting the nonlinear matter power spectrum for non-flat $\Lambda$CDM model. Section~\ref{sec:conclusion} is devoted to discussion and conclusion. In Appendix~\ref{sec:halo_model} we give justification of our method based on the halo model. Throughout this paper we use notations
$\Omega_{\rm m}$ and $\Omega_{\Lambda}$ to denote 
the density parameters for non-relativistic matter and the cosmological constant, respectively.

\section{SU estimator of $P(k)$ for non-flat $\Lambda$CDM model}
\label{sec:SU_estimator}

In this section we develop a method to compute $P(k)$ for non-flat $\Lambda$CDM model, from quantities 
for the corresponding flat $\Lambda$CDM model based on the SU approach \citep{2011JCAP...10..031B,2013PhRvD..87l3504T,2014PhRvD..89h3519L,2014PhRvD..90j3530L,2015MNRAS.448L..11W,2016JCAP...02..018L,2016PhRvD..93f3507L,2016JCAP...09..007B}.

\subsection{Preliminary}
\label{sec:preliminary}

Before going to our method we would like to introduce a motivation to use the SU approach. 
One might naively think that we can use a Taylor expansion of $P(k,z;\Omega_K)$ treating $\Omega_K$ as an expansion parameter;
$P(k,z;\Omega_K)\approx P(k,z)|_{\Omega_K=0}+(\partial P/\partial \Omega_K)|_{\Omega_K=0}\Omega_K$, where $P(k,z)|_{\Omega_K=0}$ is the power spectrum for a flat model.
However, there is no unique way to define this partial derivative operation. In particular, we have to satisfy
the identity $\Omega_K=1-(\Omega_{\rm m}+\Omega_\Lambda)$ and vary cosmological parameters other than $\Omega_K$ simultaneously.
In addition, there is ambiguity in how the time variable is matched between the flat and curved models. The simplest examples are to match the redshift, or the physical time, while these might not be optimal.
As a working example, we utilize the SU approach for connecting the power spectra for non-flat and flat $\Lambda$CDM models.

\subsection{SU approach for $P(k)$}
\label{sec:SU}

The effect of curvature ($\Omega_K$) on structure formation appears only in the late universe. In other words, 
the curvature does not affect 
structure formation in the early universe such as CMB physics (as long as the curvature parameter  is small as indicated by current observations). Hence throughout this paper we employ a model where 
structure formation in the early universe is identical. 
This is equivalent to keeping the parameters, 
\begin{align}
\left\{\omega_{\rm c}, \omega_{\rm b}, A_s, n_s\right\},
\label{eq:fixed_cosmological_parameters}
\end{align}
fixed, where $\omega_{\rm c}(\equiv \Omega_{\rm c}h^2)$ and $\omega_{\rm b}(\equiv \Omega_{\rm b}h^2)$ are the physical density parameters of CDM and baryon, respectively, and
$A_s$ and $n_s$ are the amplitude (at the pivot scale $k_{\rm pivot}=0.05\,{\rm Mpc}^{-1}$)
and the spectral tilt of the power spectrum of primordial curvature perturbations\footnote{Note that, even if we include massless and massive neutrinos, throughout this paper we fix those neutrino parameters so that 
the early universe physics remains unchanged \citep[e.g. see][for the method]{Nishimichi_2019,2021arXiv210804215B}.}.
The linear matter power spectrum is given as 
\begin{align}
P^{\rm L}(k,z)=\left(\frac{D(z)}{D(z_i)}\right)^2P^{\rm L}(k,z_i),
\label{eq:linear_ps_condition}
\end{align}
where $z_i$ is the initial redshift in the linear regime satisfying $z_i\gg 1$ 
yet well after the matter-radiation equality such that residual perturbations in radiation do not play a role
and $D(z)$ is the linear growth factor\footnote{In reality, at the typical starting redshifts of simulations, the residual radiation perturbations are not completely negligible, leaving scale-dependent corrections to the linear growth factor. 
Here we assume a situation that, as usually done when setting up the initial conditions of an $N$-body simulation, one can first evolve
the baryon and CDM perturbations until today (until the two components catch up with each other), and then trace back the ``single-fluid'' perturbation to the 
initial redshift by using the linear growth factor for a given cosmological model.}. The superscript ``L'' stands for the linear-theory quantities. 
In our method, we keep the linear power spectrum at $z_i$, $P^{\rm L}(k,z_i)$, fixed.

Let us consider a non-flat $\Lambda$CDM model, denoted as $\Omega_K$-$\Lambda$CDM, as a target model for which we want to estimate the nonlinear 
matter power spectrum at $z$ in the late universe. 
The background cosmology for this target $\KCDM$ model is specified by
\begin{align}
\mbox{$\Omega_K$-$\Lambda$CDM}: \left\{\Omega_K,\Omega_{\rm m},h\right\},
\end{align}
where $\Omega_{\rm m}$ is the density parameter of total matter (CDM plus baryon: $\Omega_{\rm m}=\Omega_{\rm c}+\Omega_{\rm b}$).
The density parameter of the cosmological constant is given by the identity,
$\Omega_K=1-(\Omega_{\rm m}+\Omega_{\Lambda})$.
The following discussion can be applied only to $\Lambda$CDM model, so we do not consider a model with 
dynamical dark energy \citep[e.g. see][for discussion on the separate universe approach for dynamical dark energy model]{2016PhRvD..94b3002H}. 

The SU approach gives a mapping between non-flat $\Lambda$CDM and 
flat $\Lambda$CDM models by assigning the degree of $\Omega_K$ in the former cosmology to the ``long-wavelength'' density fluctuation, denoted as $\delta_{\rm b}(t)$, in the latter flat 
$\Lambda$CDM model. We  call the ``fake'' flat-$\Lambda$CDM model  as $f\Lambda$CDM model.
Following \citet{2014PhRvD..89h3519L},
in the SU approach the physical matter densities in the two models are related as
\begin{align}
\bar{\rho}_{{\rm m}}(t)=\bar{\rho}_{{\rm m}f}(t)\left[1+\delta_{\rm b}(t)\right].
\end{align}
Here and throughout this paper we assume $\delta_{\rm b}(t)$ evolves according to the linear growth factor $D_f(t)$
as $\delta_{\rm b}(t)\propto D_f(t)$, and we denote quantities in the $f\Lambda$CDM model by subscript ``$f$''. 
The above equation gives 
\begin{align}
\frac{\Omega_{\rm m}h^2}{a(t)^3}=\frac{\Omega_{{\rm m}f}h_f^2}{a_f(t)^3}\left[1+\delta_{\rm b}(t)\right],
\end{align}
where we defined cosmological parameters of $f\Lambda$CDM model as $\bar{\rho}_{{\rm m}f}(a_f=1) = 
3H_{f0}^2\Omega_{{\rm m}f}/{8\pi G}$ and $H_{f0}=100~h_f~{\rm km}~{\rm s}^{-1}~{\rm Mpc}^{-1}$. 
At very high redshift in the early universe, where $|\delta_{\rm b}(t)|\ll 1$, we can find  $\Omega_{{\rm m}f}h_f^2=\Omega_{\rm m}h^2$.
This condition gives a mapping between the scale factors: 
\begin{align}
a_f(t)\left[1+\delta_{\rm b}(t)\right]^{-1/3}= a(t).
\label{eq:scale_factor_mapping}
\end{align} 
Here we stress that the mapping between quantities in $\Omega_K$-$\Lambda$CDM model and the fake flat universe should be found at the same cosmic time ($t$). The above equation allows us to find the scale 
factor $a_f(t)$ in $f\Lambda$CDM model, corresponding to $a(t)$ in  $\Omega_K$-$\Lambda$CDM model, at the same cosmic time. 
Equivalently we can find the mapping for redshift as $(1+z_f)[1+\delta_{\rm b}(z_f)]^{1/3}=1+z$.
The redshift $z_f$ in the fake universe corresponding to the target redshift $z$ can be found by solving numerically the above equation.

Once the transfer function in the early universe is fixed (Eqs.~\ref{eq:fixed_cosmological_parameters} and \ref{eq:linear_ps_condition}), 
we can take $\Omega_K$ and $h$ as the free parameters to specify the background cosmology in $\KCDM$ model. 
For a given set of $\Omega_K$ and $h$, 
we can find that the corresponding $f\Lambda$CDM model is specified by
\begin{align}
\frac{\delta_{\rm b}(t)}{D_f(t)}&=-\frac{3\Omega_K}{5\Omega_{\rm m}}\, ,\nonumber\\
 h_f &=h(1+\delta_h),
 \label{eq:hf_mapping}
\end{align}
with
\begin{align}
\delta_h\equiv (1-\Omega_K)^{1/2}-1.
\label{eq:def_delta_h}
\end{align}
Since $\delta_{\rm b}(t)\propto D_f(t)$, $\delta_{\rm b}(t)/D_f(t)$ is a constant quantity; for example,
its value at the present is specified by the first equation of Eq.~(\ref{eq:hf_mapping}). 
The condition $\Omega_{{\rm m}f}h_f^2=\Omega_{\rm m}h^2$ yields a mapping for $\Omega_{{\rm m}f}$ as
\begin{align}
 \Omega_{{\rm m}f}&= \Omega_{\rm m}(1+\delta_h)^{-2}.
 \label{eq:omegamf_mapping}
\end{align}
The flat-geometry condition for $f\Lambda$CDM model, i.e. $\Omega_{Kf}=0$ (or $\Omega_{{\rm m}f}+\Omega_{\Lambda f}=1)$, gives the following identity: 
\begin{align}
 \Omega_{\Lambda f}&=\Omega_{\Lambda}(1+\delta_h)^{-2}.
\label{eq:lambdaf_mapping}
\end{align}
In the SU approach, the effect of $\Omega_K$ on structure formation is realized by the effect of $\delta_{\rm b}$ on structure formation in a local volume in the fake flat universe.

According to the SU approach \citep{2014PhRvD..89h3519L,2014PhRvD..90j3530L}, 
the power spectrum at $z$ in $\KCDM$ model can be approximated by the power spectrum at $z_f$ in $f\Lambda$CDM model as
\begin{align}
\tilde{P}(k,z;\Omega_K)&\simeq P_f(k,z_f;\delta_{\rm b})\nonumber\\
&\hspace{-3em}\simeq \left.P_f(k,z_f)\right|_{\delta_{\rm b}=0}+\left.\frac{\partial P_f(k,z_f;\delta_{\rm b})}
{\partial \delta_{\rm b}}\right|_{\delta_{\rm b}=0}\delta_{\rm b}\nonumber\\
&\hspace{-3em}= \left.P_f(k,z_f)\right|_{\delta_{\rm b}=0}
\left[1+\left.\frac{\partial \ln P_f(k,z_f;\delta_{\rm b})}
{\partial \delta_{\rm b}}\right|_{\delta_{\rm b}=0}\delta_{\rm b}\right],
\label{eq:ps_estimator}
\end{align}
where $\delta_{\rm b}\equiv \delta_{\rm b}(z_f)$. The relation between $z$ and $z_f$ is given by Eq.~(\ref{eq:scale_factor_mapping}).
We often call $\partial P_f(k)/\partial \delta_{\rm b}$ the growth ``response'' which describes how the power spectrum at $k$ responds to the long-wavelength mode $\delta_{\rm b}$ in $f\Lambda$CDM model. 
We have put the tilde symbol $\tilde{\hspace{1em}}$ in $\tilde{P}(k,z;\Omega_K)$ to explicitly denote that $\tilde{P}$ is an ``estimator'' of the nonlinear 
matter power spectrum for $\KCDM$ model. 
Note that we need to compute these quantities at $k$ in the comoving wavenumbers of the target $\KCDM$ model, so we need not include the dilation effect, 
i.e. the mapping between comoving wavenumbers in between 
the non-flat and flat models, 
differently from the method in 
Ref.~\cite{2014PhRvD..89h3519L}.

For convenience of our discussion, we introduce the normalized response, from Eq.~(\ref{eq:ps_estimator}), as
\begin{align}
\tilde{P}(k,z;\Omega_K)\simeq 
P_f(k,z_f)\left[1+\frac{26}{21}T_{\delta_{\rm b}}(k,z_f)\delta_{\rm b}(z_f)\right],
\label{eq:ps_estimator_tb}
\end{align}
with 
\begin{align}
T_{\delta_{\rm b}}(k,z_f)\equiv \left[2\frac{\partial \ln D_f(z_f)}{\partial \delta_{\rm b}}\right]^{-1}
\left.\frac{\partial \ln P_f(k,z_f;\delta_{\rm b})}{\partial \delta_{\rm b}}\right|_{\delta_{\rm b}=0}.
\label{eq:normalized_response_db}
\end{align}
The normalized response has an asymptotic behavior of $T_{\delta_{\rm b}}\rightarrow 1$ at the linear limit 
$k\rightarrow 0$, because $P_f(k)\propto (D_f)^2P^L(k,z_i)$ in such linear regime (see Eq.~\ref{eq:linear_ps_condition}).
The coefficient, $26/21$, in the second term in the square bracket on the r.h.s. comes from the linear limit of $k\rightarrow 0$ \citep{2011JCAP...10..031B,2013PhRvD..87l3504T}.
As we will show below or discussed in Ref.~\cite{2014PhRvD..90j3530L} (around Fig.~6 in the paper), 
we propose that the power spectrum for the target $\KCDM$ model is well approximated by replacing 
the normalized response to $\delta_{\rm b}$ with the normalized response with respect to $h$ within the flat model:
\begin{align}
\tilde{P}(k,z;\Omega_K)\simeq 
P_f(k,z_f)\left[1+\frac{26}{21}T_h(k,z_f)\delta_{\rm b}(z_f)\right],
\label{eq:pk_ocdm_su_approach}
\end{align}
with 
\begin{align}
T_h(k,z_f)\equiv \left[2\frac{\partial \ln D_f(z_f)}{\partial h_f}\right]^{-1}
\frac{\partial \ln P_f(k,z_f)}{\partial h_f},
\label{eq:normalize_response_h}
\end{align}
where the partial derivative $\partial/\partial h_f$ is the derivative with respect to $h_f$, while keeping 
the other cosmological parameters (Eq.~\ref{eq:fixed_cosmological_parameters})
fixed to their fiducial values; more explicitly we vary $h$ with 
keeping $\Omega_{\rm m}h^2$ fixed, and accordingly we have to change $\Omega_{\rm m}$ (and 
$\Omega_{\Lambda}$ from the identity $\Omega_\Lambda=1-\Omega_{\rm m}$ for flat models). 
Here we defined the normalized response satisfying $T_h(k)\rightarrow 1$ at $k\rightarrow 0$.
If $T_{\delta_{\rm b}}(k,z_f)\simeq T_h(k,z_f)$ for an input set of $k$ and $z_f$ as we will show below, 
we can use Eq.~(\ref{eq:pk_ocdm_su_approach}) to approximate the nonlinear matter power spectrum at $z$ for the target 
$\KCDM$ model.

\subsection{Linear limit}
\label{sec:spherical_collapse}

The SU picture has an analogy with the spherical collapse model \citep{1967PThPh..37..831T,1972ApJ...176....1G,2012PhRvD..85f3521I}, where a spherical 
top-hat over- or under-density fluctuation is embedded into the FRW homogeneous background
and then the time-evolution of the top-hat interior density can be fully tracked up to the fully nonlinear regime.
As described in \citet{2015MNRAS.448L..11W}, we can find a mapping between the full growth factor of such a spherical tophat density and the 
linearly-extrapolated density fluctuation $\delta_{\rm b}$ up to the full order of $\delta_{\rm b}$. 
In the SU setup this is equivalent to expressing the growth factor of density fluctuations in a local volume with $\delta_{\rm b}$,
denoted as $\tilde{D}_f(z_f;\delta_{\rm b})$, 
in terms of the growth factor in the background of the fake universe \citep[from Eq.~22 in Ref.][]{2015MNRAS.448L..11W}:
\begin{align}
\tilde{D}_f(z_f;\delta_{\rm b})\simeq D_f(z_f)
\left[1+\frac{13}{21}\delta_{\rm b}+\frac{71}{189}\delta_{\rm b}^2+\frac{29609}{130977}\delta_{\rm b}^3
\right],
\label{eq:linear_growth_db}
\end{align}
where $\delta_{\rm b}=\delta_{\rm b}(z_f)$.
Comparison of Eqs.~(\ref{eq:ps_estimator_tb}) and (\ref{eq:linear_growth_db}) clarifies that 
the expression of Eq.~(\ref{eq:ps_estimator_tb}) corresponds to the linear-order expansion of the linear growth factor 
in terms of $\delta_{\rm b}$, because $P_f(k,z_f;\delta_{\rm b})\propto \tilde{D}_f(z_f,\delta_{\rm b})^2$ at the linear limit. 
The coefficient $26/21$ on the r.h.s. in Eq.~(\ref{eq:pk_ocdm_su_approach}) comes from the first-order 
expansion of the linear growth factor in the above equation: $(\tilde{D}_f/D_f)^2\simeq 1+(26/21)\delta_{\rm b}$.
Because we know the exact mapping between the linear growth factors in the $\KCDM$ and $f\Lambda$CDM models, we can fully account for the 
mapping at the linear limit. We will later include this linear-limit correction.

\begin{figure}
	\includegraphics[width=0.49\textwidth]{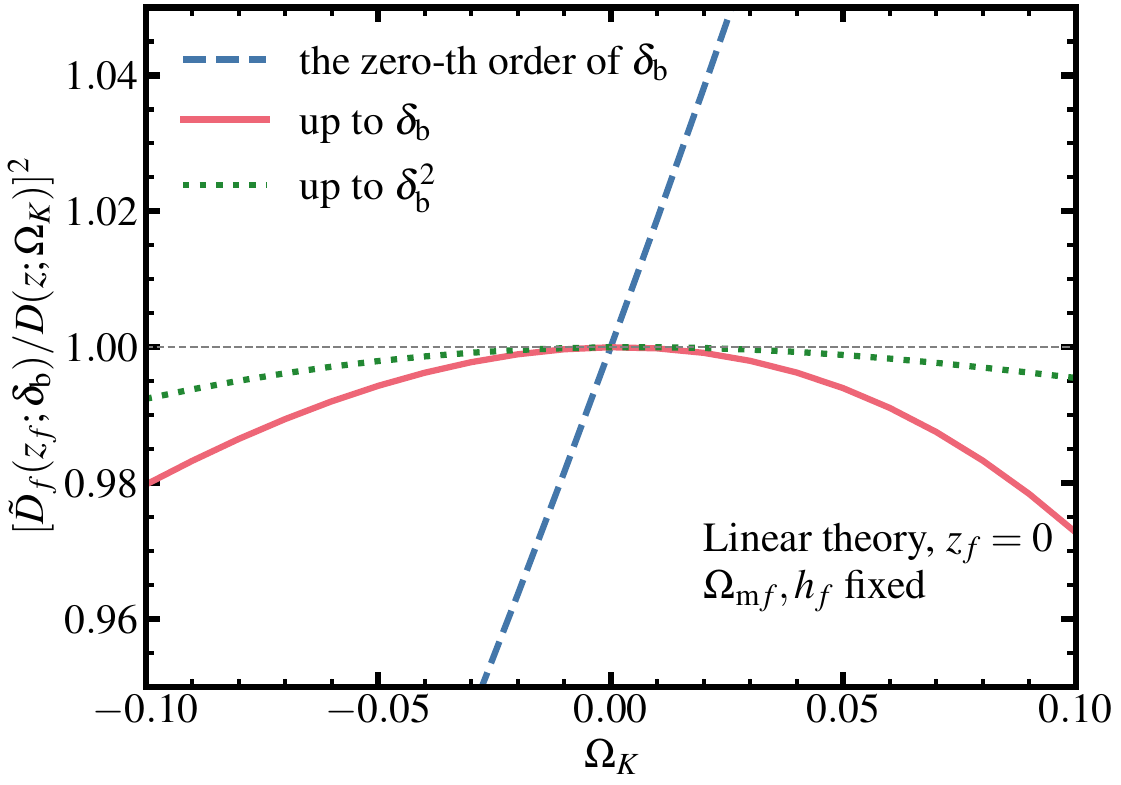}
	\caption{An accuracy of the approximation that  gives 
    the growth factor    for 
    $\KCDM$ model  in terms of the growth factor for the corresponding 
    flat $\Lambda$CDM model and the Taylor expansion of 
    $\delta_{\rm b}$ in the SU approach (Eq.~\ref{eq:linear_growth_db}). Here 
    $\delta_{\rm b}$ is related to the curvature parameter $\Omega_K$, in the $x$-axis, of each $\KCDM$ model,
    via Eq.~(\ref{eq:hf_mapping}). Note that we consider $z_f=0$ 
    and fixed the other cosmological parameters 
    $\Omega_{{\rm m}f}$ and $h_f$ as $\Omega_{{\rm m}f}=0.3156$ and $h_f=0.6727$ in the flat $\Lambda$CDM model. 
    Here we assess $[\tilde{D}_f(z_f;\delta_{\rm b})/D(z;\Omega_K)]^2$, where $D(z;\Omega_K)$ is the true growth factor for each $\KCDM$ model, because the ratio corresponds to the linear limit of the approximation of matter power 
    spectrum we study in this paper.   
    The dashed, solid and dotted curves denote the results for the approximations obtained when including the zeroth-,
    first- or second-order expansion of $\delta_{\rm b}$ in Eq.~(\ref{eq:linear_growth_db}).
   	}
	\label{fig:linearlimit}
\end{figure}

Fig.~\ref{fig:linearlimit} shows the accuracy of the approximation of the growth factor, 
$[\tilde{D}_f(z_f;\delta_{\rm b})/D(z;\Omega_K)]^2$ 
when truncated at some finite order in $\delta_{\rm b}$, as a function of the input $\Omega_K$ in the $x$-axis, where $D(z;\Omega_K)$ is the true growth factor 
for $\KCDM$ model. 
The value of $\delta_{\rm b}$ is specified by the input $\Omega_K$ in 
the $x$-axis, from Eq.~(\ref{eq:hf_mapping}).
The dashed, solid and dotted curves show the ratio when including 
only the zeroth term, or up to the 1st or 2nd term, respectively, in the square bracket of the r.h.s. 
of Eq.~(\ref{eq:linear_growth_db}). 
The 1st-order expansion (solid curve) corresponds to the approximation of the power spectrum, 
$\tilde{P}(k,z;\Omega_K)/P(k,z;\Omega_K)$ (Eq.~\ref{eq:ps_estimator_tb}) at linear limit ($k\rightarrow 0$).
Encouragingly, the 1st-order approximation (solid curve) is accurate to within about 2\% in the fractional amplitude for the range 
of $|\Omega_K|<0.1$, which is very broad compared to the current constraint, %$|\Omega_K|\lesssim O(0.1)$.
$|\Omega_K|<0.1 \, (2 \sigma \, {\rm level})$ from the {\it Planck} CMB data alone \cite{planck_collaboration_2020}.
Note that, if we do not take into account the mapping of redshift $(z_f\leftrightarrow z)$ or we forcibly use the power spectrum at $z$ in the fake universe, the accuracy of the 1st-order approximation is significantly degraded. 
We also note that the results are similar for other redshifts, but have better accuracy with the increase of redshift.

\subsection{Summary: Estimator of $P(k)$ for non-flat $\Lambda$CDM}

By combining Eqs.~(\ref{eq:pk_ocdm_su_approach}) and (\ref{eq:linear_growth_db}), 
we propose the following approximation to compute the nonlinear matter power spectrum at $z$ for $\KCDM$ model that is specified by the parameters $(\Omega_K,\Omega_{\rm m},h)$: 
\begin{align}
\tilde{P}(k,z)&\simeq P_f(k,z_f)\left(\frac{D(z)}{D_f(z_f)}\right)^2 \nonumber\\
&\hspace{4em}\times\left[1+\frac{26}{21}\left\{T_h(k,z_f)-1\right\}\delta_{\rm b}\!(z_f) \right],
\label{eq:pk_kcdm_estimator}
\end{align}
with 
\begin{align}
&\delta_{\rm b}(z_f)=-D_f(z_f)\frac{3\Omega_K}{5\Omega_{\rm m}},\nonumber\\
&(1+z_f)\left[1+\delta_{\rm b}(z_f)\right]^{1/3}=1+z, \nonumber\\
&T_h(k,z_f)=\left[2\frac{\partial \ln D_f(z_f)}{\partial h_f}\right]^{-1}\frac{\partial \ln P_f(k,z_f)}{\partial h_f}.
\end{align}
Note that the parameters ($h_f,\Omega_{{\rm m}f},\Omega_{\Lambda f}$) for $f\Lambda$CDM model are given by Eqs.~(\ref{eq:hf_mapping}),
(\ref{eq:omegamf_mapping}) and (\ref{eq:lambdaf_mapping}),
and the other cosmological parameters to specify the transfer function and the primordial perturbations ($\omega_{\rm c},\omega_{\rm b},A_s,n_s$) are kept fixed in the 
$\KCDM$ and $f\Lambda$CDM models. 

We employed the modification of Eq.~(\ref{eq:pk_kcdm_estimator}) from Eq.~(\ref{eq:pk_ocdm_su_approach}) to fully take into account the 
modification in the linear growth factor up to the full order of $\delta_{\rm b}$; Eq.~(\ref{eq:pk_kcdm_estimator}), by design, reproduces the underlying true power spectrum for $\KCDM$ model at the linear limit, i.e. $\tilde{P}(k,z)=P(k,z)$ at $k\rightarrow 0$ (also from Eq.~\ref{eq:linear_ps_condition}).

All the terms on the r.h.s. of Eq.~(\ref{eq:pk_kcdm_estimator}), except for 
$D(z)$,  
are given by quantities for the flat-geometry $f\Lambda$CDM model, which are specified by the cosmological parameters of $\KCDM$ model. 
That is, if Eq.~(\ref{eq:pk_kcdm_estimator}) is a good approximation, we can evaluate the nonlinear matter 
power spectrum for an arbitrary $\KCDM$ model from the quantities for the counterpart flat model in the SU approach. 
For example, if we use the fitting formulae or emulators
of nonlinear matter power spectrum calibrated for flat
$\Lambda$CDM cosmologies, we can compute the nonlinear matter power spectrum for the target $\KCDM$ model.
This would be a useful approximation, and we will below give validation of our method, and quantify the accuracy of our method.

\section{Simulation Data}
\label{sec:simulations}

%%%%%%%%%%%%%%%%%%%%%%%%%%%%%%%%%%%%%%%%%%%%%%%%%%%
\begin{table*}
\caption{Details of $N$-body simulations for different cosmological models. 
The columns ``$\Omega_K$'' and ``$h$'' give their values of the curvature parameter and Hubble parameter
that are  
employed in each simulation, while we fix 
other cosmological parameters $\{\omega_{\rm c},\omega_{\rm b}, A_s, n_s\}$, which are needed to specify the linear power spectrum for the initial conditions, to the values for the fiducial 
{\it Planck} cosmology (see text for details). 
$\Omega_{\rm m}$ and $\Omega_\Lambda$ are specified by a given set of $\Omega_K$ and $h$, 
because we keep 
$\Omega_{\rm m}h^2$ fixed and $\Omega_{\Lambda}=1-\Omega_{\rm m}-\Omega_{\rm K}$.
The column ``$N_{\rm real}$'' denotes the number of realizations, with different initial seeds, used for each model. 
The column ``Angulo-Pontzen'' gives whether or not we employ the ``paired-and-fixed'' method in \cite{2016MNRAS.462L...1A} to reduce the sample variance effect in small $k$ bins for the power spectrum measurement: we adopt the method for ``Yes'', while not for ``No''. For the paired-and-fixed method, it uses the paired (2) simulations by design (see text for details). 
The column ``redshift ($z$)''
gives the redshifts of simulation outputs: for $\KCDM$ model, we properly choose the redshifts corresponding to the same cosmic time for each of redshifts, $z=\{0.0,0.549,1.025,1.476\}$ in  the ``fiducial'' model in the SU approach (see around Eq.~\ref{eq:scale_factor_mapping} in Section~\ref{sec:SU}).  
All the simulations are done in the fixed comoving box size without $h$ in its units, i.e. $L\simeq1.49\,{\rm Gpc}$ (corresponding to $1\,h_f^{-1}{\rm Gpc}$ for the fiducial model)
and with the same particle number, i.e. $N_p=2048^3$. 
\label{tab:simulations}
}
\begin{center}
\begin{tabular}{l|ccccl}\hline\hline 
Name & $\Omega_K$ & $h$ &  $N_{\rm real}$  & Angulo-Pontzen  & redshift ($z$) \\ \hline 
flat (fiducial) & 0 & $0.6727$ &  2 & Yes. & $\{0.0, 0.549, 1.025, 1.476\}$ \\ \hline
$\Omega_K$-$\Lambda$CDM1 & 0.00663 & 0.6749 & 10  & No. &$\{-0.0033,  0.544,  1.018,  1.467\}$ \\
& $-0.00672$ & 0.6705  & 10  &No. &$\{0.0033, 0.554, 1.031, 1.484\}$ \\ \hline 
$\Omega_K$-$\Lambda$CDM2 & 0.05 & 0.6902  & 2 & Yes. &$\{-0.027,   0.518,  0.992,  1.443\}$\\
& $-0.05$ & 0.6565&  2 & Yes. & $\{0.023 , 0.576,  1.053, 1.505\}$ \\ \hline
$\Omega_K$-$\Lambda$CDM3 & 0.1 & 0.7091  &2 & Yes. & $\{-0.059,    0.482,  0.955,  1.405\}$ \\
& $-0.1$ &0.6414  &2 & Yes. &  $\{0.043 ,0.600,  1.079,  1.531\}$\\ \hline 
$\delta h$-$\Lambda$CDM & 0 & 0.6927 & 10  & No. &$\{0.0, 0.549, 1.025, 1.476\}$\\ 
& 0& 0.6527 & 10  & No. &$\{0.0, 0.549, 1.025, 1.476\}$ \\ \hline\hline
\end{tabular}\\
\end{center}
\end{table*}
%%%%%%%%%%%%%%%%%%%%%%%%%%%%%%%%%%%%%%%%%%%%%%%%%%%

\subsection{$N$-body simulations}
\label{sec:n-body}

To validate our method (Eq.~\ref{eq:pk_kcdm_estimator}), we use cosmological $N$-body simulations.
Our simulations follow the method in \citet{Nishimichi_2019}, and we here give a brief summary of the simulations used in this paper. 

We use {\tt Gadget-2} \citep{gadget2} to carry out $N$-body simulation for a given cosmological model. 
The initial conditions are set up at redshift $z_i = 59$
using the second-order Lagrangian perturbation theory 
\citep{scoccimarro98,crocce06b} implemented by \citet{nishimichi09}
and then parallelized in \citet{Valageas11a}. 
We use the public code {\tt CAMB} \citep{camb} to compute 
the transfer function for a given model, which is used 
to compute the input linear power spectrum.
For all simulations in this paper, 
we use the same simulation box size in Gpc (i.e., without $h$ in the units) and the same number of particles:  
$L = 1\,h_f^{-1}\mathrm{Gpc}\simeq 1.49\,{\rm Gpc}$ (without $h$ in units)
and $N_p=2048^3$, which correspond to the particle Nyquist wavenumber, $k = 6.4 ~ h_f \rm{Mpc}^{-1}$.
In the following we will show the results at wavenumbers smaller than this Nyquist wavenumber.

In this paper we use simulations for 5 different cosmological models, denoted as ``fiducial'' flat $\Lambda$CDM, 
``$\KCDM$1'', ``$\KCDM$2'', ``$\KCDM$3'', and ``$\delta h$-$\Lambda$CDM'' models, respectively, as given 
in Table~\ref{tab:simulations}. Here the cosmological parameters for the ``fiducial'' model are chosen to be consistent with those
for the {\it Planck} 2015 best-fit cosmology~\cite{planck-collaboration:2015fj}. The cosmological parameters for each of the non-flat cosmological models 
are chosen so that it has the fiducial $\Lambda$CDM model as the ``fake'' flat $\Lambda$CDM model in the SU approach. 
We use paired simulations for ``$\KCDM$1'' model to compute the power spectrum response 
with respect to $\delta_{\rm b}$ ($T_{\delta_{\rm b}}$), where the curvature parameters are specified by 
$\delta_{\rm b}=\pm 0.01$ at $z_f=0$. The ``$\delta h$-$\Lambda$CDM'' model is for computing the response with respect to $h$ ($T_h$): 
here, we chose a step size of $\delta h= \pm 0.02$ for the numerical derivative.
Note that the setup of these simulations is designed to compute the ``growth'' response by taking the numerical derivative
at fixed comoving wavenumbers $k$ \citep[see Section~IIIB in Ref.][]{2014PhRvD..89h3519L}. 
We also use the simulations for non-flat $\Lambda$CDM models with 
$\Omega_K=\pm 0.05$ or $\pm 0.1$, named as ``$\KCDM$2'' and ``$\KCDM$3'',  
to assess how our method can approximate the matter power spectrum for non-flat models.

Table~\ref{tab:simulations} gives the values of $\Omega_K$ and $h$, and we use the fixed values of other cosmological parameters, given as $(\omega_{\rm c},\omega_{\rm b},A_s, n_s)=(0.1198,0.02225,2.2065 \times 10^{-9}, 0.9645)$, which 
specify the transfer function and the primordial power spectrum, or equivalently the linear matter power spectrum\footnote{Note that we 
also include the effect of massive neutrinos on the linear matter power spectrum, assuming 
$\Omega_\nu h^2=0.00064$ corresponding to $m_{\nu,{\rm tot}}=0.06\,{\rm eV}$, the lower limit inferred from 
the terrestrial experiments \citep[see Ref.][for details]{Nishimichi_2019}. Hence the physical density parameter of total matter is 
$\Omega_{\rm m}h^2=\omega_{\rm c}+\omega_{\rm b}+\omega_\nu$.}. 
Note that $\Omega_{\rm m}$ and
$\Omega_\Lambda$ are specified 
by a given set of the parameters for each model: $\Omega_{\rm m}=\Omega_{{\rm m}f}h_f^2/h^2$ and $\Omega_K=1-(\Omega_{\rm m}+\Omega_{\Lambda})$. 
For each model, we use the outputs at 4 redshifts, $z_f\simeq 0, 0.55, 1.03$ and $1.48$. Since the ``fiducial''
flat $\Lambda$CDM model is the fake flat model in the SU method, each redshift 
for the fiducial flat model corresponds to a slightly different redshift in each non-flat model, which 
is computed from Eq.~(\ref{eq:scale_factor_mapping}).

Note that all the $N$-body simulations for different cosmological models are designed to have 
the fixed mass resolution, $m_p\simeq 1.52\times 
10^{10}~M_\odot$ (in units without $h$).
Hence the comoving mass density in the $N$-body box 
is kept fixed: 
$\bar{\rho}_{\rm m0}=N_{p}m_{p}/V_{\rm com}\simeq 3.96
\times 10^{11}~M_\odot{\rm Mpc}^{-3}$.
We utilize this fact to define a sample of halos in 
the same mass bins, in units of $M_\odot$,  
for all the cosmological models. This makes it easier to compute the response of halo mass function with respect
to $\delta_{\rm b}$ or $h$, which is used to study the power spectrum responses based on the 
halo model (see Appendix~\ref{sec:halo_model}).

Furthermore, we use simulations that are run using the ``paired-and-fixed'' method in \citet{2016MNRAS.462L...1A}, where 
the initial density field in each Fourier mode is generated from the fixed amplitude of the power spectrum 
$\sqrt{P(k)}$ and the paired simulations with reverse phases, i.e. $\delta_{\bk}$ and $-\delta_{\bk}$, are run. The mean power spectrum of the paired runs fairly well reproduces 
the ensemble average of many realizations even in the nonlinear regime \cite{2016MNRAS.462L...1A,2018ApJ...867..137V}. The paired-and-fixed simulations allow us to significantly reduce the sample variance in the power spectrum estimation. 
The column ``Angulo-Pontzen'' in Table~\ref{tab:simulations} denotes whether we use the paired-and-fixed simulations. For the paired-and-fixed simulations, ``2'' on the column $N_{\rm real}$ denotes one pair of the pared-and-fixed simulations.

\subsection{Measurements of power spectrum and growth response}
\label{sec:sim_powerspectrum}

To calculate the power spectrum from each simulation output, we assign the particles on $2048^3$ grids using 
the cloud-in-cells (CIC) method \citep{hockney81} to obtain the density field. After performing the Fourier transform, 
we correct for the window function of CIC following the method described in \citet{Takahashi12}.
In addition, to evaluate the power spectrum at small scales accurately, we fold the particle positions
into a smaller box by replacing $\bold{x} \rightarrow \bold{x}\%(L/10^n)$, where the operation $a\% b$ stands for
the reminder of the division of $a$ by $b$. This procedure leads to effectively $10^n$ times higher resolution.
Here we adopt $n = 0,1$. We use the density fluctuation $\delta_{\bold{k}}$ up to half the Nyquist frequency determined by 
the box size $L/10^n$ with the grid number, and we will show the results at wavenumbers smaller than $k = 6.4~h_f{\mathrm{Mpc}^{-1}}$. 

Since we use the fixed box size and the same particle number, we 
use the same $k$ binning to estimate the average of $|\delta_{\bf k}|^2$ in each $k$ bin to estimate the band power. 
We then use the two-side numerical derivative method to compute the power spectrum responses. 
To reduce statistical stochasticity (or sample variance), we employ the same initial seeds as those for the ``fiducial'' model. The column ``$N_{\rm real}$'' 
in Table~\ref{tab:simulations} denotes the number of realizations for paired simulations, where each pair uses the same initial seeds. For $\KCDM$1 and 
$\delta h$-$\Lambda$CDM models, we further run 9 paired simulations to estimate the statistical scatters; 
hence we use 10 paired simulations in total to estimate the power spectrum responses at each redshift, 
$T_{\delta_{\rm b}}(k, z_f)$ and $T_h(k, z_f)$.

\section{Results}
\label{sec:results}

\subsection{Power spectrum responses}
\label{sec:pk_response}

In Fig.~\ref{fig:Tdeltab,h} we study the normalized growth responses of matter power spectrum, $T_{\delta_{\rm b}}(k)$ 
and $T_h(k)$, at the four redshifts, 
which are computed from the $N$-body simulations for $\KCDM1$ and $\delta h$-$\Lambda$CDM models, respectively, 
in Table~\ref{tab:simulations}. It is clear that the approximate identity of 
$T_{\delta_{\rm b}}\approx T_h$ holds over the range of scales and for all the redshifts.
To be more precise, the two responses agree with each other to within $2~(16)\%$ in the fractional amplitudes for $k \lesssim 1~(6.4)~h_f{\rm Mpc}^{-1}$.
Our results confirm the result of \citet{2014PhRvD..90j3530L} (see Fig.~6 in the paper).
However, a closer look of Fig.~\ref{fig:Tdeltab,h} reveals a slight discrepancy at $k\gtrsim 1\,h_f{\rm Mpc}^{-1}$. 
As we showed in Appendix~\ref{sec:halo_model}, the responses at these small scales are mainly from modifications in the mass density profiles of halos.
Hence we conclude that the identity of $T_{\delta_{\rm b}}\approx T_h$ is not exact, but approximately valid
for models around the fiducial $\Lambda$CDM models we consider.

\begin{figure*}
	\includegraphics[width=0.45\textwidth]{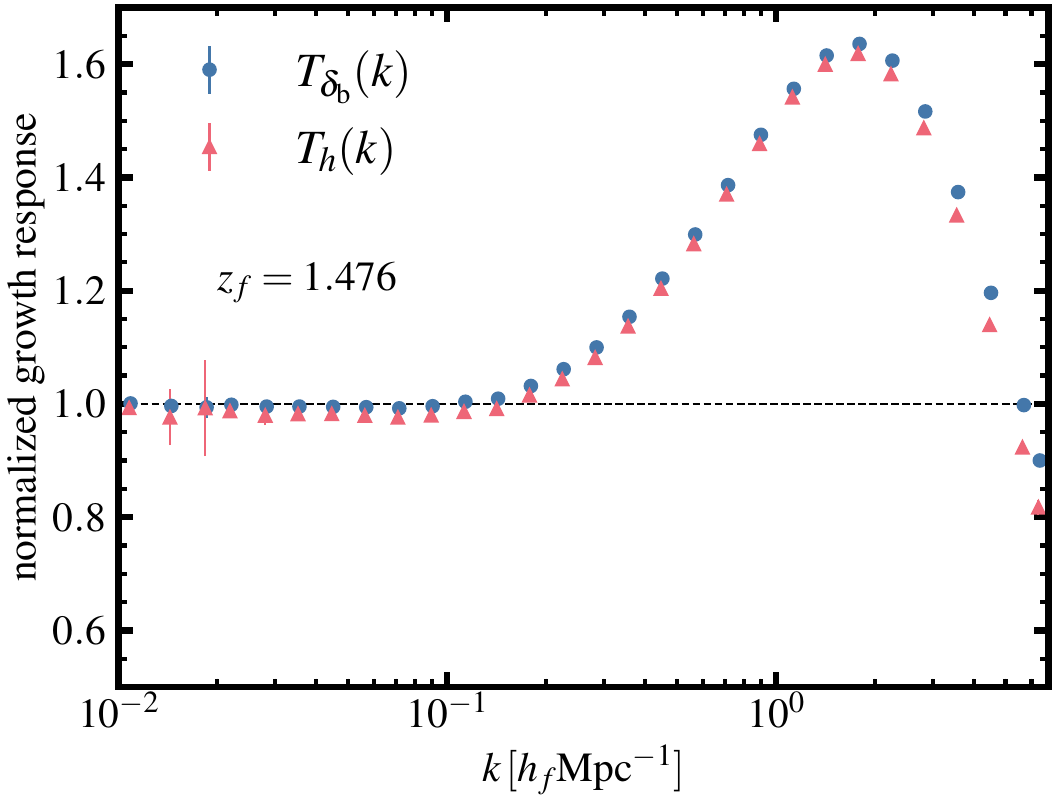}
	\includegraphics[width=0.45\textwidth]{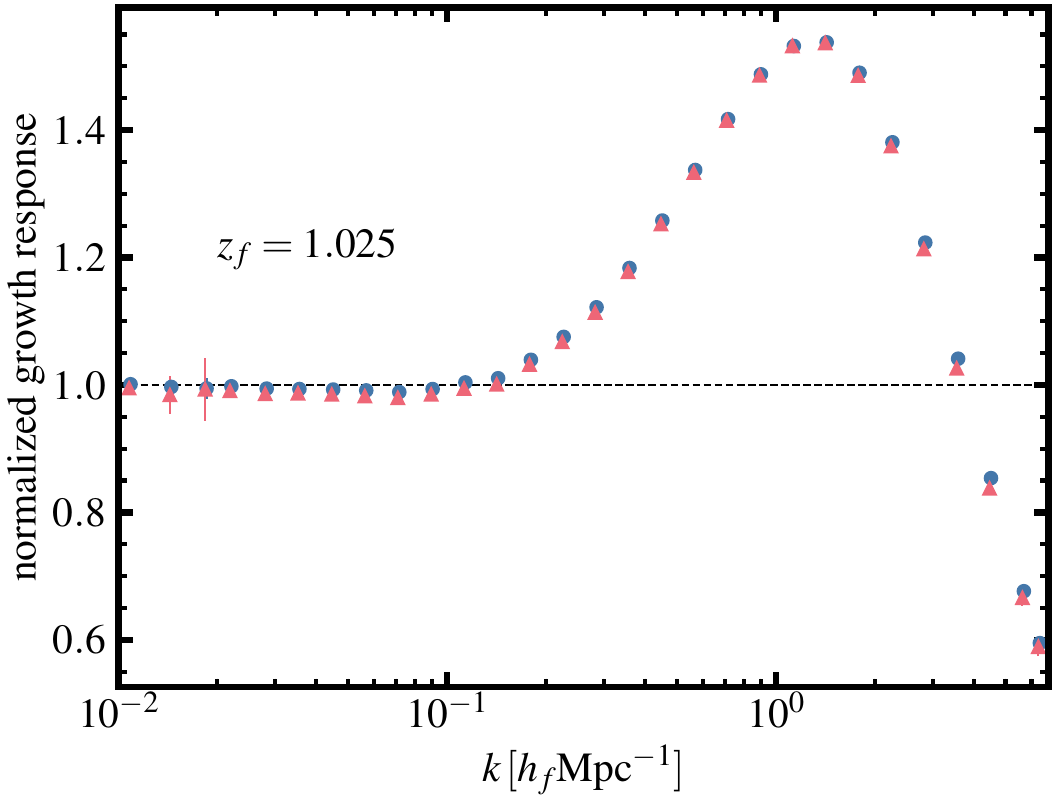}
	\includegraphics[width=0.45\textwidth]{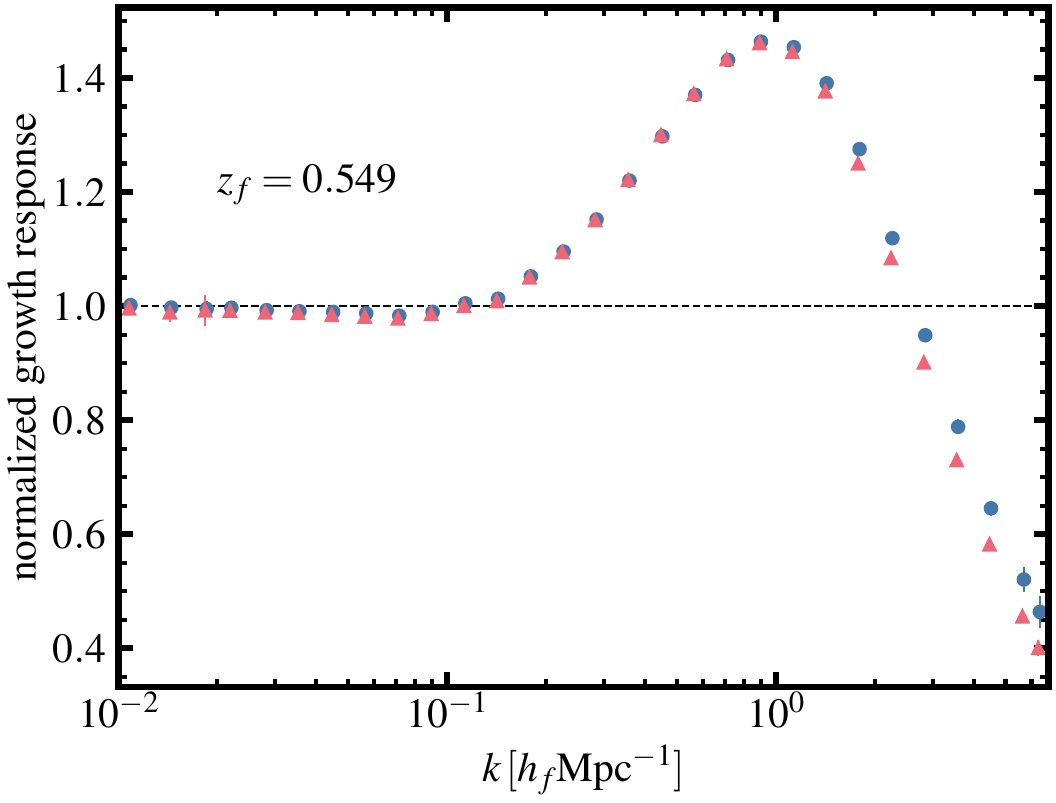}
	\includegraphics[width=0.45\textwidth]{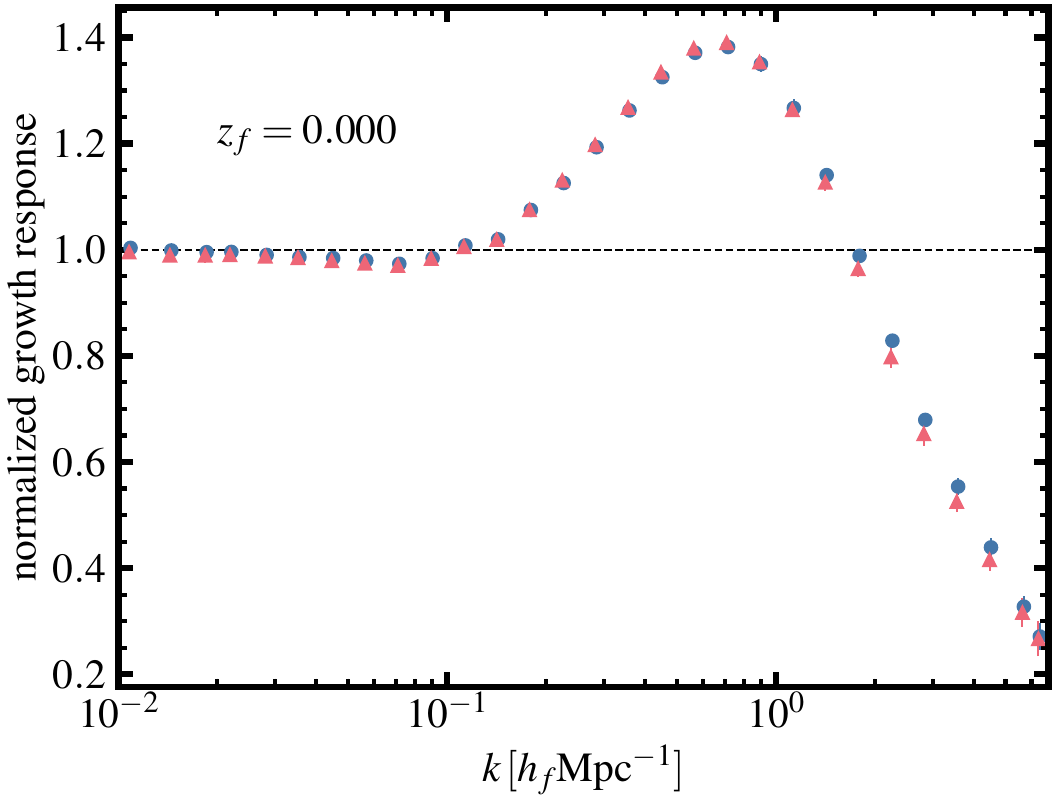}
	\caption{Normalized growth response of matter power spectrum with respect to $\delta_{\rm b}$ and $h$, 
    $T_{\delta_{\rm b}}(k)$ and $T_h(k)$, 
    at 
    the four redshifts as denoted by the legend in each panel. The horizontal line 
    denotes the linear limit: $T_{\delta_{\rm b}},T_h=1$.
    We use the 10 paired simulations for $\KCDM1$ and $\delta h$-$\Lambda$CDM models in Table~\ref{tab:simulations} to compute these responses. 
    The circle or triangle symbols denote the mean of $T_{\delta_{\rm b}}$ or $T_h$ 
    in each $k$ bin, and the error bars (although not visible in some $k$ bins) denote the statistical errors for simulation box with side length $L=1\,h_f^{-1}{\rm Gpc}$, which 
    are estimated from the standard deviations among the 10 paired simulations. Note that the range of $y$-axis is different in different panels.
  }
	\label{fig:Tdeltab,h}
\end{figure*}
\begin{figure}
    \includegraphics[width=0.45\textwidth]{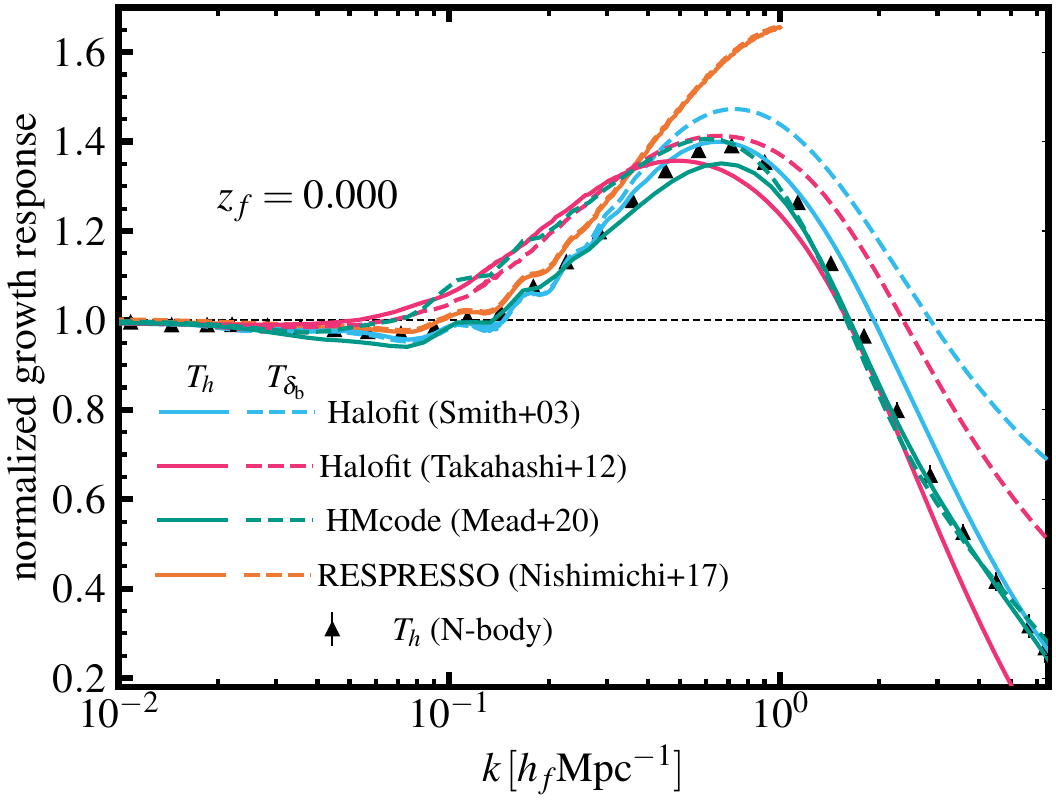}
    \caption{Comparison of the simulation results of $T_h$ and $T_{\delta_{\rm b}}$ at $z_f=0$
    with the predictions computed from the public codes of nonlinear matter power spectrum: 
    ``Smith+03 {\tt Halofit}'' \citep{Smith03}, 
    ``Takahashi+12 {\tt Halofit}'' \citep{Takahashi12}, ``Mead+20 {\tt HMcode}'' \citep{2021MNRAS.502.1401M},
    and ``Nishimichi+17 {\tt RESPRESSO}'' \citep{2017PhRvD..96l3515N}.
    Here we used the public codes to compute the normalized responses
    from numerical derivative of the power spectrum predictions with models varying 
    $\Omega_K$ or $h$ (see text for details).
    The solid and dashed lines show the model predictions for $T_h$ and 
    $T_{\delta_{\rm b}}$, respectively. The triangle symbols are the same as in Fig.~\ref{fig:Tdeltab,h}. 
    We omit the simulation result for $T_{\delta_{\rm b}}$ as it is very close to the triangle symbols, to avoid crowdedness in the figure.
    }
    \label{fig:Tdeltab,h,Halofit}
\end{figure}
\begin{figure}
    \includegraphics[width=0.45\textwidth]{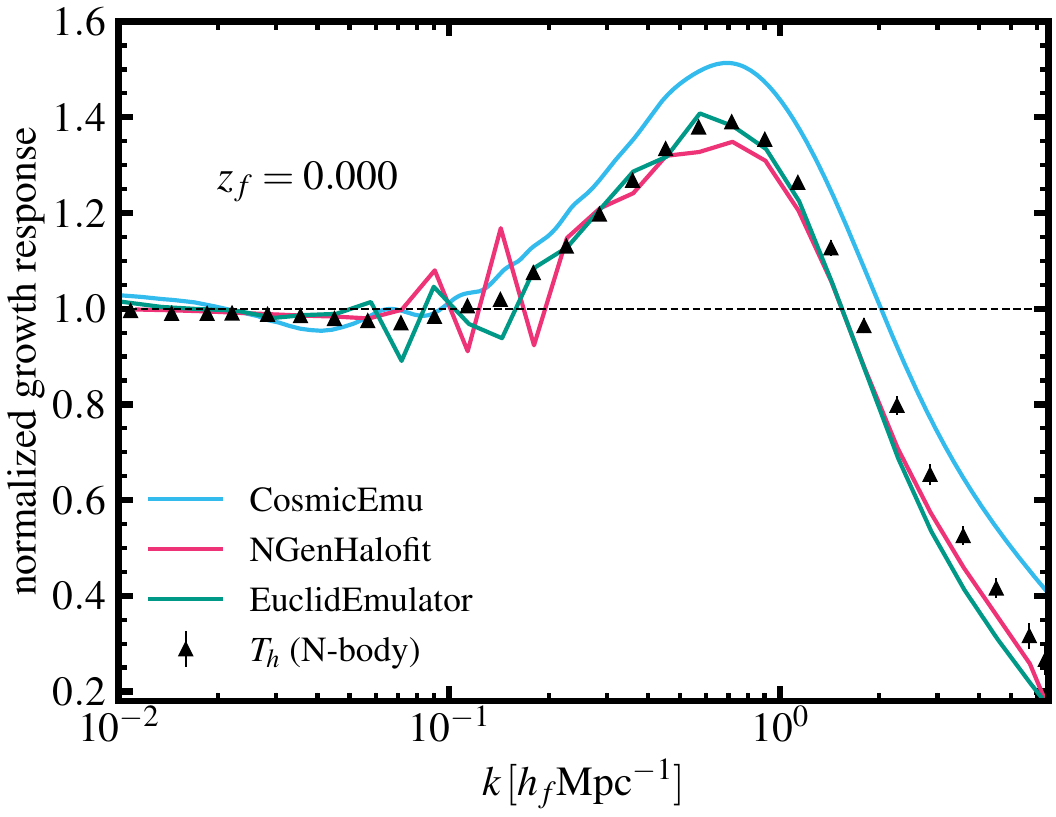}
    \caption{Similar to the previous figure, but this figure compares the simulation result for $T_h$ at $z_f=0$ with those computed from the public emulator of the nonlinear matter power spectrum:
    {\tt CosmicEmu} \citep{MiraTitan1}, {\tt NGenHalofit} \citep{2019MNRAS.486.1448S} and {\tt EuclidEmulator} \citep{2021MNRAS.505.2840E}. 
    These emulators are designed to compute the power spectrum for flat-geometry CDM cosmologies, so here we compare the results for $T_h$. 
    }
    \label{fig:Tdeltab,h,emulators}
\end{figure}

Since $\delta_{\rm b}$ and $h$ are varied at fixed initial power spectrum, their impact on the power spectrum in the linear regime at a fixed comoving
scale comes solely from changing the linear growth factor $D$. 
Since we normalize the response to account for this linear dependence on $D$, the data points converge to unity at the low-$k$ limit by design.

For the quasi nonlinear regime, the perturbation theory of structure formation predicts that 
the higher-order loop corrections to the power spectrum are 
well approximated by a separable form in terms of time and scale: an exact result for the Einstein-de Sitter (EdS) cosmology, which is usually generalized to $\Lambda$CDM cosmology by replacing the scale factor $a$ with the linear growth factor $D$.
Possible corrections to this arising from the non-separability is known to have a weak dependence 
on $\Omega_{\rm m}(z)$, and this can be usually ignored in the modeling of mildly nonlinear regime \citep{bernardeau02,takahashi:2008lr,2014phrvd..90b3546n,2016PhRvD..94b3504T}.
Under this approximation, the nonlinear power spectrum is fully specified by its linear counterpart evaluated at the same redshift.
Therefore the perturbation theory also predicts that $T_{\delta_{\rm b}}\approx T_h$ 
should be valid due to the matched shape of the linear power spectrum. 

The agreement $T_{\delta_{\rm b}}\approx T_h$ in the nonlinear regime suggests that nonlinear matter power spectrum 
is approximately given by a functional form of the input linear power spectrum, $\Delta^2_{\rm NL}(k)=F_{\rm NL}[\Delta^2_{\rm L}(k)]$
($\Delta^2\equiv k^3P(k)/2\pi^2)$, as implied 
by the stable clustering ansatz for a CDM model \citep{1991ApJ...374L...1H,1994MNRAS.267.1020P,1995MNRAS.276L..25J,jain:1996lr,1996MNRAS.280L..19P}. If the ansatz holds, the identity 
$T_{\delta_{\rm b}}=T_h$ holds exactly.
In Appendix~\ref{sec:halo_model}, we also show that the approximation $T_{\delta_{\rm b}}\approx T_h$ can be found from the halo model 
picture, which is derived using the growth responses of the halo mass function and the halo mass density profile to $\delta_{\rm b}$ and $h$ that are estimated from $N$-body simulations. Thus the results of Fig.~\ref{fig:Tdeltab,h} 
suggest that the stable clustering ansatz is approximately valid. 

Before proceeding, we comment on the normalized growth response to the primordial power spectrum amplitude $A_{\rm s}$: $T_{A_s}(k,z_f) \equiv \left[2{\partial \ln D_f(z_f)}/{\partial A_s}\right]^{-1} {\partial \ln P_f(k,z_f)}/{\partial A_s}$. 
A change in $A_{\rm s}$ does not alter 
the shape of the linear matter power spectrum.
If the stable clustering is exact, one would expect $T_{A_s}=T_{\delta_{\rm b}}$. 
However, as shown in Fig.~6 of \citet{2014PhRvD..90j3530L} 
\citep[also see][]{2015JCAP...08..042W}, $T_{A_{\rm s}}$ shows a sizable discrepancy from $T_{\delta_{\rm b}}$ (or 
$T_h$) at $k\gtrsim 0.1\,h_f{\rm Mpc}^{-1}$. 
This means that a change in $A_{\rm s}$ leads to a larger 
change in the transition scale ($k_{\rm NL}$) between the linear and nonlinear regimes, or a 
larger change in the halo profile (e.g., the halo concentration). 
Hence, we again stress that $T_{\delta_{\rm b}}(k)\approx T_h(k)$ is an approximate identity
around the $\Lambda$CDM model.  

In Fig.~\ref{fig:Tdeltab,h,Halofit} we assess the accuracy of the publicly-available fitting formula of $P(k)$
for predicting the normalized growth response.
Here we employ the two versions of {\tt Halofit} in \citet{Smith03} (Smith+03) and \citet{Takahashi12} (Takahashi+12), respectively, 
and the {\tt HMcode} in \citet{2021MNRAS.502.1401M} (Mead+20). All the fitting formulae are primarily functional of the linear 
power spectrum at the target redshift (although each formula includes terms that have an extra dependence on cosmological parameters).
Among these fitting formulae, only Smith+03 {\tt Halofit} was calibrated against $N$-body simulations for models including non-flat CDM model. 
Note that we here compute $T_{\delta_{\rm b}}(k)$ from Eq.~(\ref{eq:ps_estimator_tb}) based on the SU method; 
we compute $\tilde{P}(k,z)$ for varied $\KCDM$ models
assuming that the fitting formula is valid for non-flat cosmologies, respectively, 
and then compute $T_{\delta_{\rm b}}(k)$ from numerical derivative.
Here we adopt $\delta_{\rm b}=\pm 0.01$ at $z_f = 0$.  
On the other hand, for $T_h(k)$, 
we vary only $h$ with keeping the linear matter power spectrum fixed (keeping $\Omega_{\rm m}h^2$ fixed as discussed around Table~\ref{tab:simulations}) in  flat models, 
and then 
compute $T_h(k)$  
from numerical derivative of the fitting formula predictions, where
we adopt variations of $\delta h=\pm 0.02$.
The figure shows that none of the fitting formulae reproduces the approximate identity of $T_{\delta_{\rm b}}\approx T_h$ in nonlinear regime at the level that we see in the responses measured from $N$-body simulations \citep[also see Ref.][for the similar discussion]{2020MNRAS.492.5226R}. 
This implies that the fitting formulae have a degraded accuracy for non-flat cosmologies in the 
nonlinear regime, 
because the response
to $\delta_{\rm b}$ is equivalent to the dependence of $P(k)$ on $\Omega_K$.
Nevertheless it is 
intriguing to find that all the fitting formulae give a closer prediction to the simulation result for
the response to $h$, 
than that to $\delta_{\rm b}$.
In particular, the result for {\tt HMcode} appears to be promising, including a better agreement over
the transition scales between the linear and nonlinear regimes. 

In Fig.~\ref{fig:Tdeltab,h,Halofit}, we also show the response computed using {\tt RESPRESSO} \citep{2017PhRvD..96l3515N}. 
It reconstructs the nonlinear power spectrum from 
an input linear power spectrum at a target redshift for a target cosmological model based on the perturbation theory motivated method 
starting from a nonlinear matter power spectrum at a fiducial cosmology measured from $N$-body simulations.
The difference in the nonlinear power spectra in $k$ bin between the two cosmologies is computed by summing up the contributions from the
band power of the linear power spectrum in each $q$ bin.
To do this, the diagrams in the perturbative expansion relevant to the response of the nonlinear power spectrum to the linear counterpart are precomputed at different values of $\Omega_\mathrm{m}$, and this lookup table is inferred along a path between the fiducial and the target cosmology.
Note that {\tt RESPRESSO} outputs the predictions up to $k \sim 1~h\rm{Mpc}^{-1}$.
Since {\tt RESPRESSO} needs only the linear power spectrum of the target cosmology as input, its prediction is unique for models with the same linear power spectrum (e.g., two models with different values of $A_\mathrm{s}$, but evaluated at different redshifts to match the normalization of the linear power spectrum), even when the growth history is different.
Hence $T_h$ and $T_{\delta_{\rm b}}$ computed by {\tt RESPRESSO} are indistinguishable except for the slight difference due to numerical accuracy.
However, since {\tt RESPRESSO} is motivated by the perturbation theory,
these predictions begin to 
differ from the responses measured from $N$-body simulations around $k \sim 0.2~h \mathrm{Mpc}^{-1}$,
which corresponds to the scale where the perturbation theory fails.
This is similar to the results from Ref.~\cite{2015JCAP...08..042W} (see Fig.~2 in the paper),
where the perturbation theory (1-loop) predictions were used to compute the response up to quasi nonlinear regime.
These results indicate that the perturbation theory motivated model fails to predict the responses in the nonlinear regime. 
That is, as we discussed above, the nonlinear power spectrum has other dependencies besides 
the linear power spectrum (also see Appendix~\ref{sec:halo_model} for the similar discussion).

In Fig.~\ref{fig:Tdeltab,h,emulators} we use the 
the publicly-available emulators of $P(k)$, built for flat cosmologies, to assess accuracy for predicting the response to $h$.
Here we used {\tt CosmicEmu \citep{MiraTitan1}}, {\tt NGenHalofit \citep{2019MNRAS.486.1448S}}, and {\tt EuclidEmulator \citep{2021MNRAS.505.2840E}}.
The later two emulators fairly well reproduce the simulation results, although a jagged feature in the numerical derivative is seen, probably due to a  $k$-binning issue (or interpolation issue) in the output.

\subsection{Accuracy of the approximation of $P(k)$ for non-flat $\Lambda$CDM}
\label{sec:accuracy}

\begin{figure*}
    \includegraphics[width=0.45\textwidth]{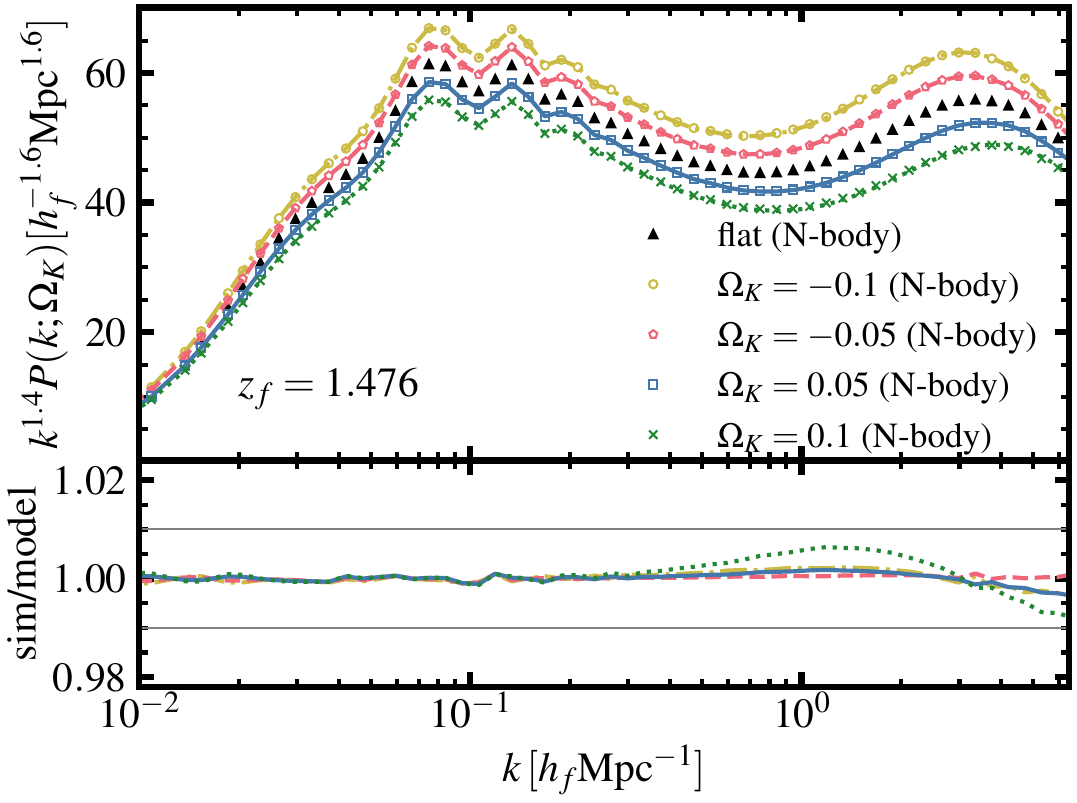}
    \includegraphics[width=0.45\textwidth]{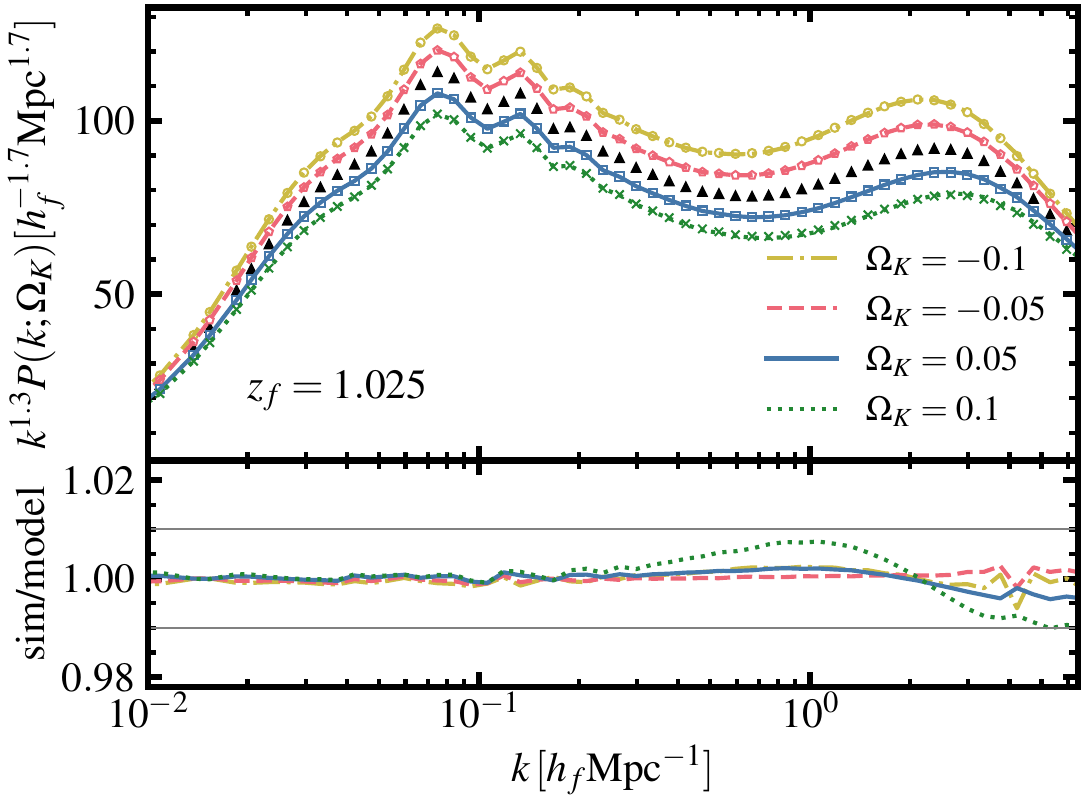}
    \includegraphics[width=0.45\textwidth]{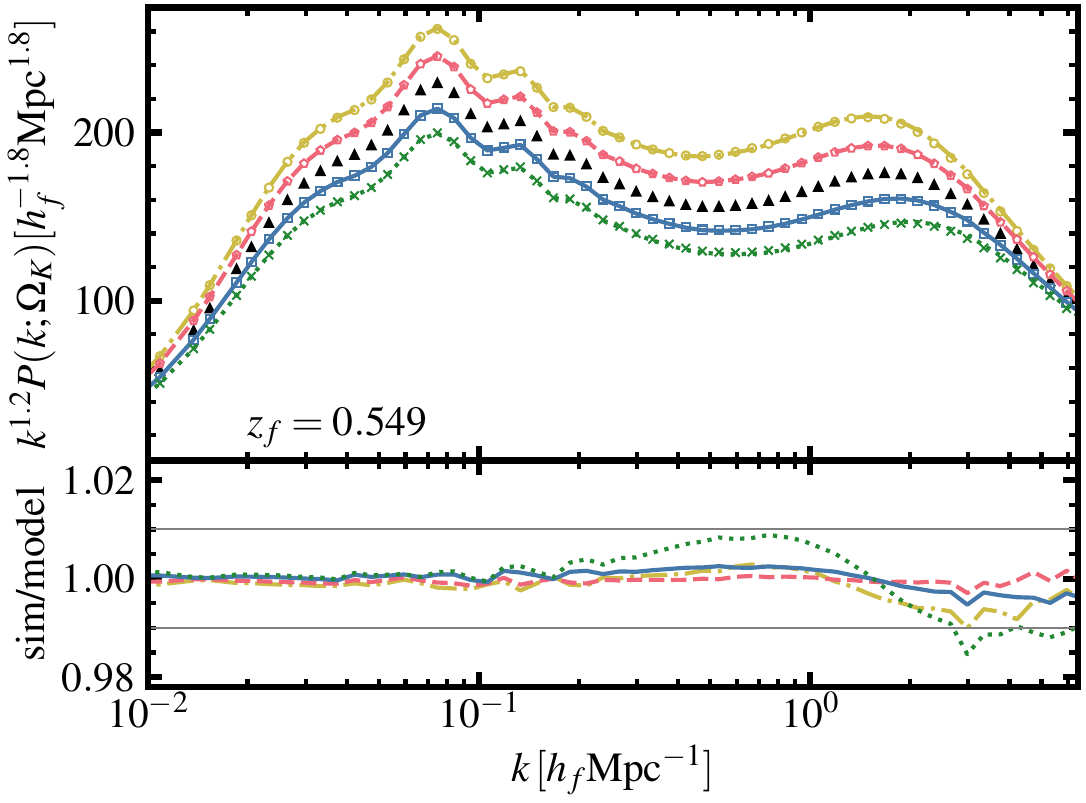}
    \includegraphics[width=0.45\textwidth]{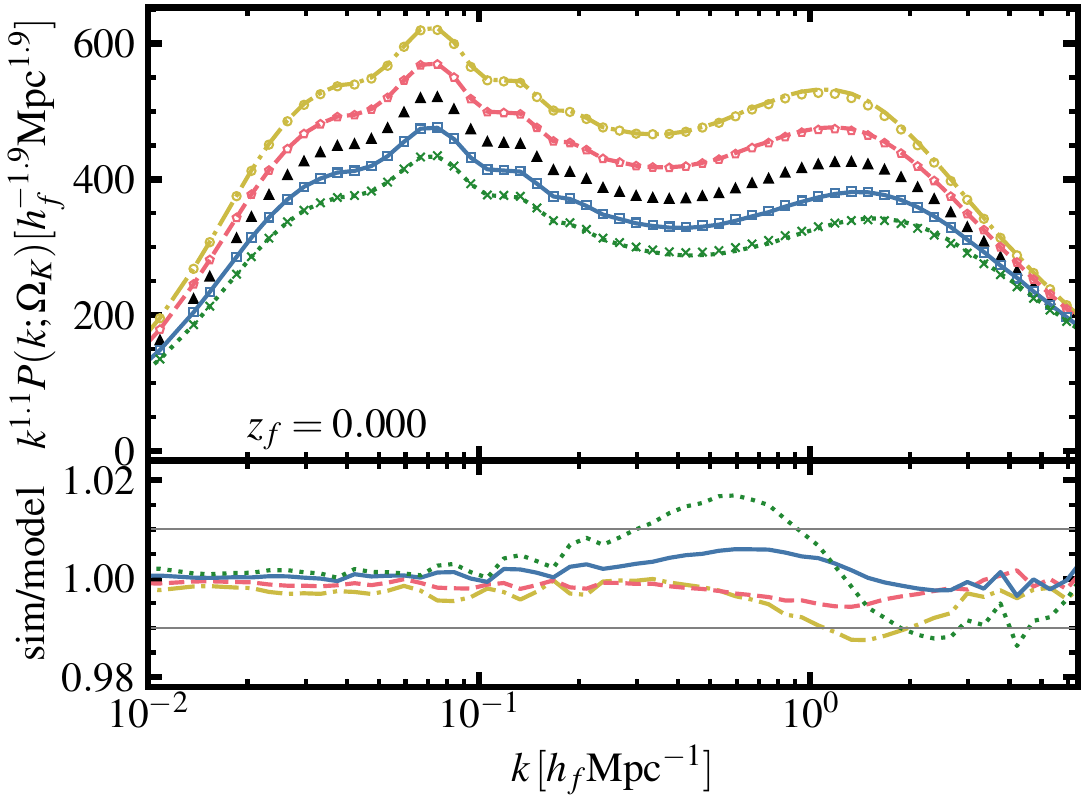}
    \caption{An assessment of the accuracy of our method (Eq.~\ref{eq:pk_kcdm_estimator}) for predicting 
    $P(k)$ for non-flat $\Lambda$CDM models. The different symbols in each panel denote $P(k)$,
    directly estimated from $N$-body simulations for non-flat 
    models with $\Omega_K=\pm 0.05$ and $\pm0.1$ ($\KCDM2$ and $\KCDM3$ models in Table~\ref{tab:simulations}), at each of the 
    four redshifts, while the lines denote the results from our method, $\tilde{P}(k)$ in Eq.~(\ref{eq:pk_kcdm_estimator}).
    Note that we used the simulation results for $P_f(k,z_f)$ and $T_h(k,z_f)$ in Eq.~(\ref{eq:pk_kcdm_estimator}). 
    For the simulation results we used the ``paired-and-fixed''
    method in \citet{2016MNRAS.462L...1A} to reduce the stochasticity. For comparison we also show the simulation result for the flat fiducial simulation by triangle symbols. For illustrative purpose we show $k^\alpha P(k,z)$ where the power index $\alpha$ for each redshift output is chosen in that the $y$-range is narrower in the linear scale. 
    The lower plot in each panel shows the ratio between the simulation result and our method. The horizontal lines denote $\pm 1\%$ fractional accuracy. 
    }
    \label{fig:OmegaKs}
\end{figure*}
\begin{figure*}
    \includegraphics[width=0.45\textwidth]{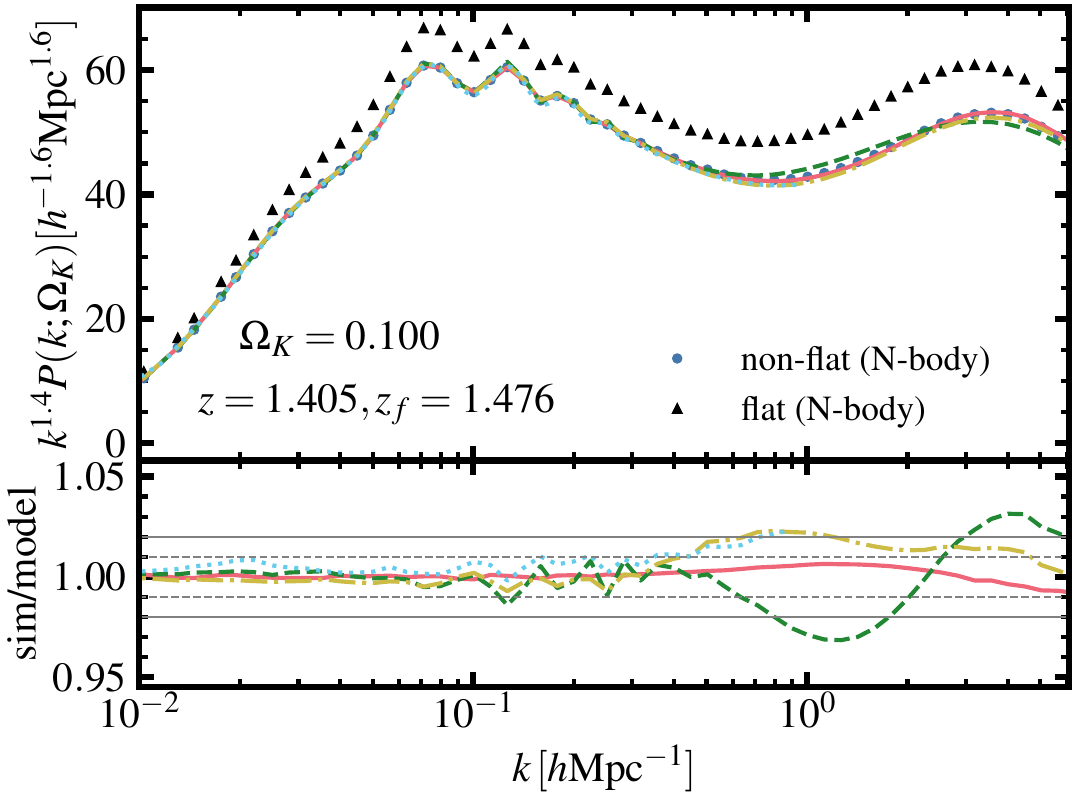}
    \includegraphics[width=0.45\textwidth]{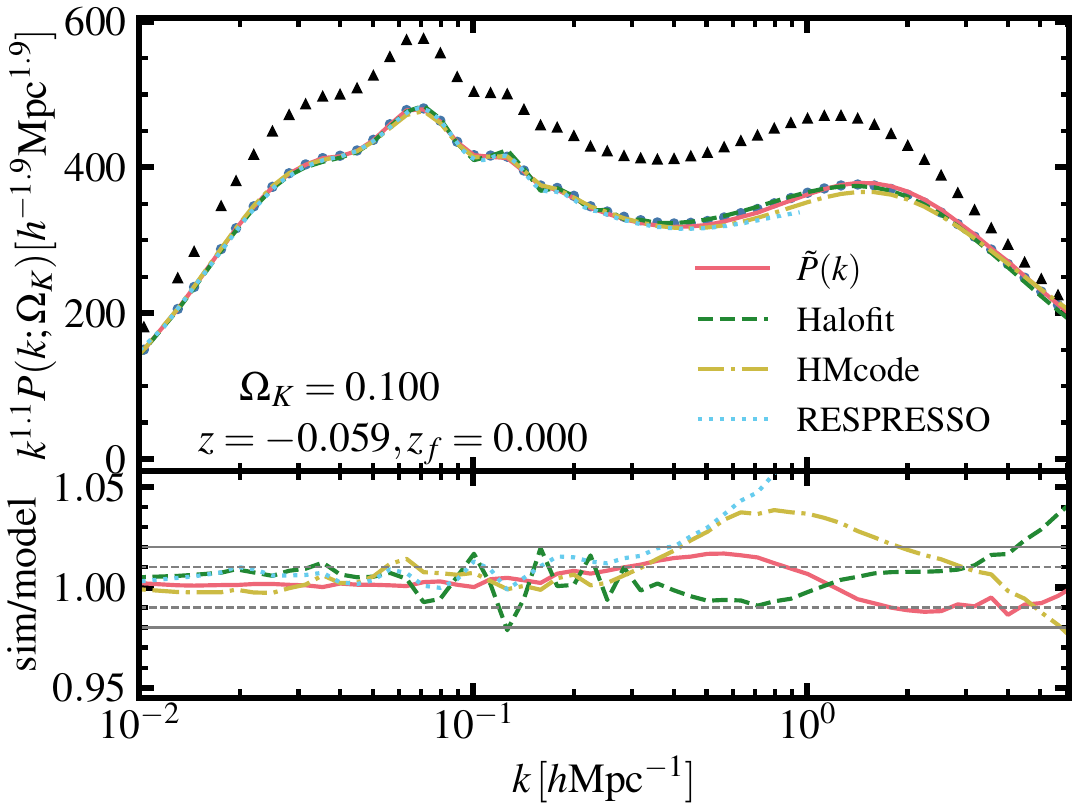}
    \includegraphics[width=0.45\textwidth]{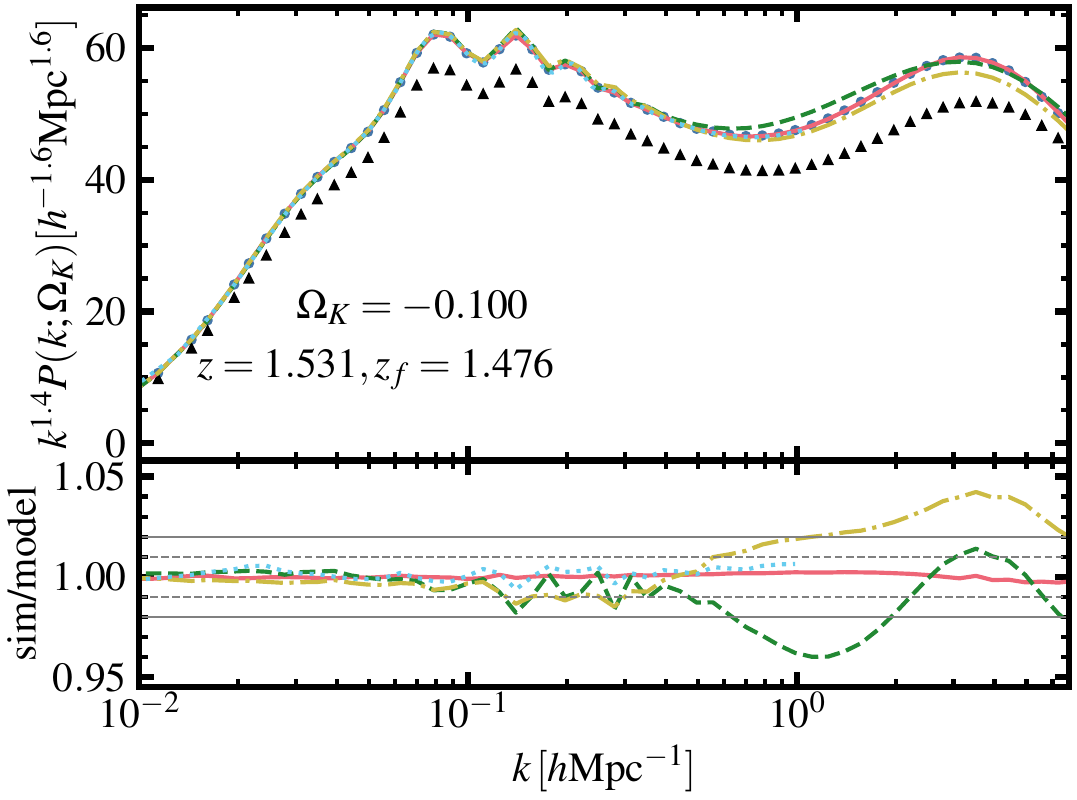}
    \includegraphics[width=0.45\textwidth]{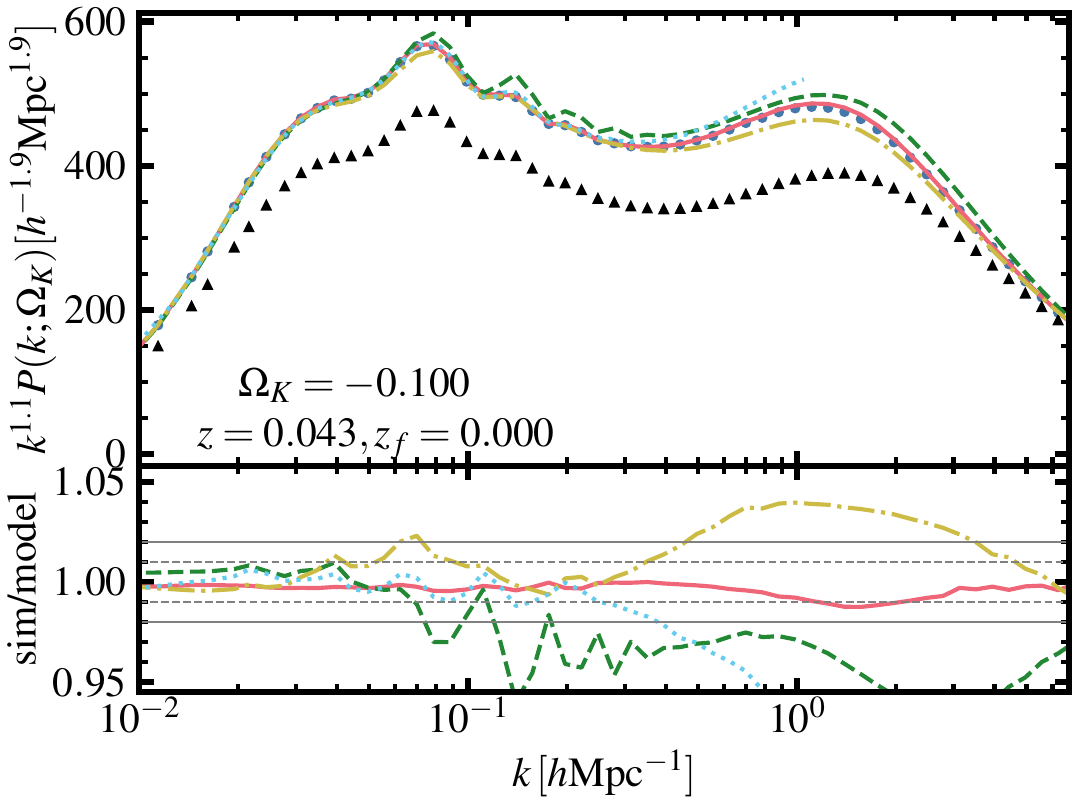}
    \caption{Similar to the previous figure, but this figure compares the nonlinear matter power spectra that are 
    computed from our method and the public codes, for the $\KCDM3$ model with $\Omega_K=\pm 0.1$. Here 
    we consider {\tt Halofit} \citep{Takahashi12}, {\tt HMcode} \citep{2021MNRAS.502.1401M} and {\tt RESPRESSO}
    \citep{2017PhRvD..96l3515N}. The solid line in each panel is the same as in the previous figure, while
    the other lines denote the results computed from the codes, where 
    we used the direct predictions for non-flat models. 
    The horizontal dashed and solid lines denote $\pm 1\%$ and $\pm 2\%$ fractional accuracy, respectively. 
    }
    \label{fig:Halofit_OK=+-0.1}
\end{figure*}
In this section we assess an accuracy of the approximation (Eq.~\ref{eq:pk_kcdm_estimator}) to 
evaluate $P(k)$ for $\KCDM$ model, by comparing the predictions based on the method with 
the power spectra directly measured from $N$-body simulations for $\KCDM$ model.

The data points in Fig.~\ref{fig:OmegaKs}
show $P(k)$ estimated from the simulations, in 
Table~\ref{tab:simulations}, for $\KCDM$ models with 
$\Omega_K = \pm \{0.05, 0.1\}$, at each of the four redshifts. 
The curves in each panel show the predictions computed based on Eq.~(\ref{eq:pk_kcdm_estimator}), where 
we used the normalized response $T_h$, computed from the simulations (the results in 
Fig.~\ref{fig:Tdeltab,h}), and 
the power spectrum $P_f(k,z_f)$ computed from  the $f\Lambda$CDM simulation (the triangle symbols).
The figure shows that the estimator reproduces the simulation result at an accuracy better than $\sim 1\%$ in the amplitude over the
wide range of wavenumbers, except for $\sim 2\%$ accuracy 
for $\Omega_K=0.1$ at $k\sim 1\,h_f{\rm Mpc}^{-1}$ that 
corresponds to the largest $\delta_{\rm b}$. 
The $1\%$ accuracy at $k\sim 1\,h_f^{-1}{\rm Mpc}$
roughly meets requirements on $P(k)$ for upcoming weak lensing surveys \citep{Huterer:2004tr}.

In Fig.~\ref{fig:Halofit_OK=+-0.1}, we compare the performance of 
the estimator predictions (Eq.~\ref{eq:pk_kcdm_estimator}) 
with public codes for each cosmological model. 
Here we consider two fitting formulae, {\tt Halofit} in Takahashi et al. \citep{Takahashi12} and {\tt HMcode} \citep{2021MNRAS.502.1401M}. In addition, we consider {\tt RESPRESSO} \citep{2017PhRvD..96l3515N}. %based on the power-spectrum total-response method, which reconstructs the power spectrum for a target model starting from that at a nearby fiducial model precomputed using $N$-body simulations.
Although these models are calibrated under $\Omega_K=0$, we use their direct predictions for non-flat cosmologies using an extrapolation to $\Omega_K \ne 0$. 
While the accuracy of {\tt Halofit}, {\tt HMcode} and {\tt RESPRESSO} vary with the redshift, the estimator prediction displays a better performance 
than the public codes especially in the nonlinear regime.

\begin{figure}
    \includegraphics[width=0.45\textwidth]{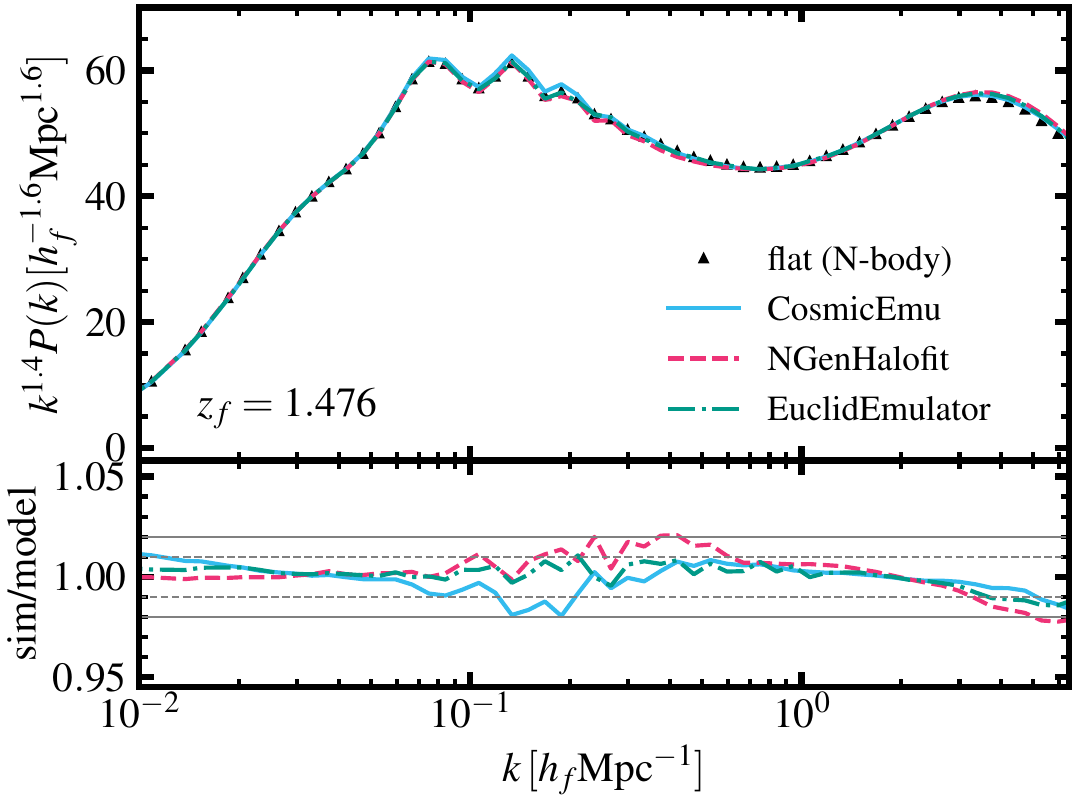}
    \includegraphics[width=0.45\textwidth]{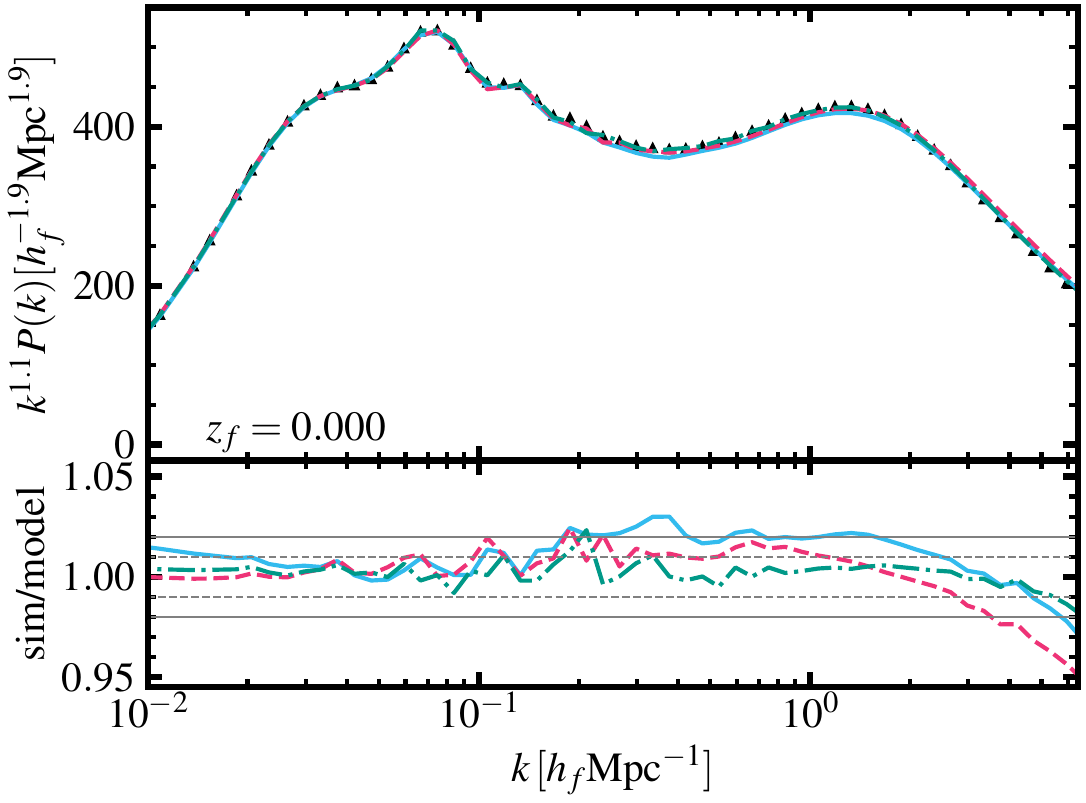}
    \caption{Comparison of nonlinear power spectra for flat cosmology at $z_f=1.476$ (top) $z_f=0$ (bottom) computed from our simulations and the public emulators.
  Here we consider {\tt CosmicEmu \citep{MiraTitan2}},
  {\tt NGenHalofit \citep{2019MNRAS.486.1448S}}, and {\tt EuclidEmulator \citep{2021MNRAS.505.2840E}}.
  The horizontal dashed and solid lines denote $\pm 1\%$ and $\pm 2\%$ fractional accuracy, respectively. 
  }
    \label{fig:Emulator_flat}
\end{figure}
\begin{figure}
    \includegraphics[width=0.45\textwidth]{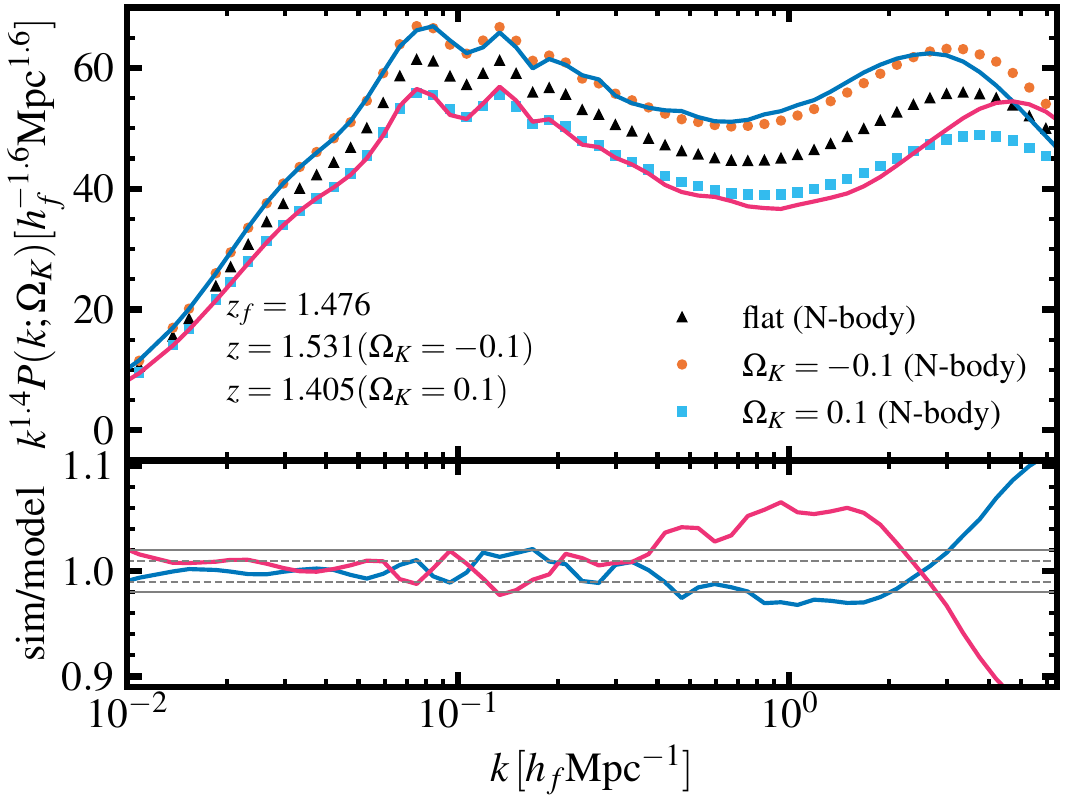}
    \includegraphics[width=0.45\textwidth]{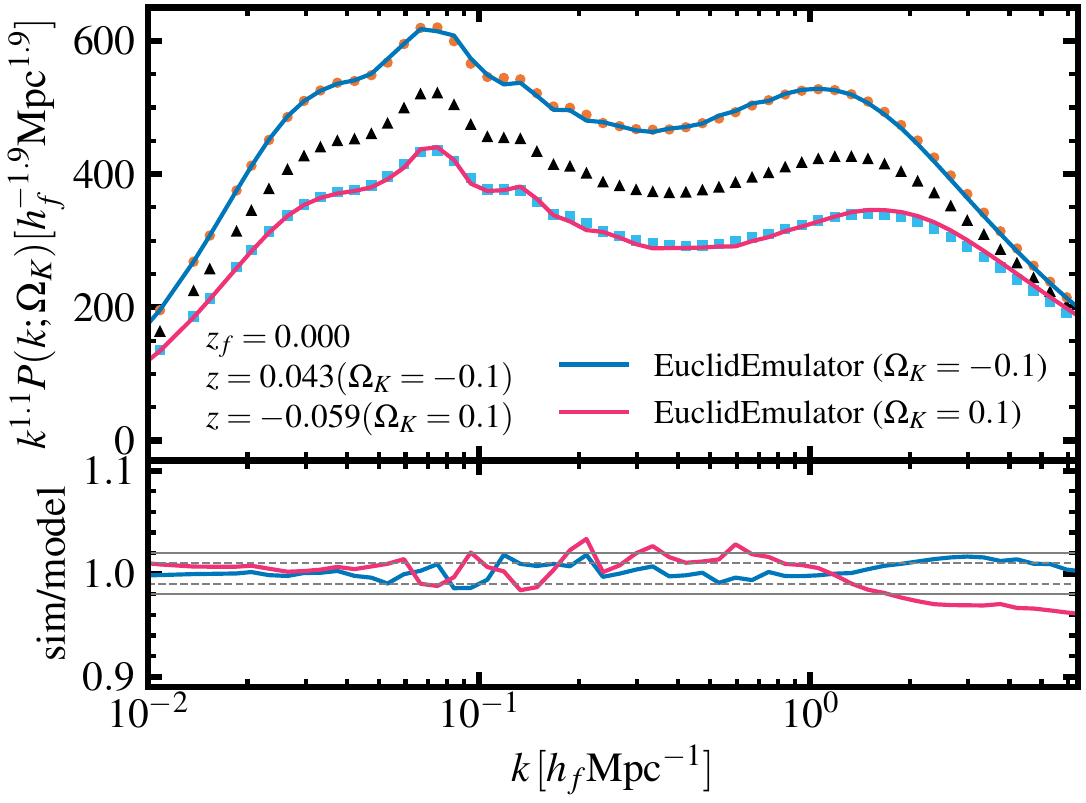}
    \caption{Performance of our method (Eq. 17) for predicting the nonlinear matter power spectrum for non-flat $\Lambda$CDM model with $\Omega_K = \pm 0.1$, at $z_f = 1.476$ (top) and $z_f=0$ (bottom). 
  Here we use {\tt EuclidEmulator} \citep{2021MNRAS.505.2840E} predictions for $P_f(k,z_f)$ and $T_h(k,z_f)$ in Eq.~(\ref{eq:pk_kcdm_estimator}).
  The horizontal dashed and solid lines denote $\pm 1\%$ and $\pm 2\%$ fractional accuracy, respectively. 
  }
    \label{fig:EmulatorExtension}
\end{figure}
Now we study the accuracy of the emulators of $P(k)$, calibrated only for flat cosmologies, to predict 
$P(k)$ for non-flat models, based on our method (Eq.~\ref{eq:pk_kcdm_estimator}). Before going to the result, in Fig.~\ref{fig:Emulator_flat}, we assess the performance of the emulators by 
comparing the predictions with the simulation results for the fiducial flat model.
All the emulators reproduce the simulation results to within $\sim 2\%$ up to $k\sim 1\,h_f {\rm Mpc}^{-1}$.

In Fig.~\ref{fig:EmulatorExtension} we apply our method to 
the {\tt EuclidEmulator} for predicting $P(k)$ for the non-flat models with $\Omega_K=\pm0.1$,
compared to the simulation results, where we use the emulator to compute $T_h(k,z_f)$ and $P_f(k,z_f)$
and then plug those into Eq.~(\ref{eq:pk_kcdm_estimator}) to obtain $\tilde{P}(k,z)$ for the 
target $\KCDM$ model. The predictions by {\tt EuclidEmulator} are in good agreement with the simulations at $z_f = 0$
to within $\sim 2\%$ even for the relatively large $|\Omega_K|$ models, 
while the accuracy is degraded especially at 
high-k for $z_f \simeq 1.5$. The nice agreement for $z_f=0$ is encouraging, and the accuracy of the emulator for the high redshift might need to be further studied.

\section{Discussion and Conclusion}
\label{sec:conclusion}

In this paper we have developed an approximate method to compute the nonlinear matter power spectrum, 
$P(k)$, for a ``non-flat'' $\Lambda$CDM model, from quantities computed for the counterpart flat $\Lambda$CDM model, base on the separate universe (SU) method. 
To do this, we need to employ a specific mapping of the cosmological parameters and redshifts
between the non-flat and flat models, with keeping the initial power spectrum fixed. In addition 
we utilized the fact that the normalized response of $P(k)$ to the long-wavelength fluctuation mode $\delta_{\rm b}$ in the flat model
is well approximated by the normalized response to $h$ for the flat model, which was validated by using the $N$-body simulations.  
We showed that our method (Eq.~\ref{eq:pk_kcdm_estimator}) enables to compute $P(k)$ for non-flat 
models with $|\Omega_K|\le 0.1$, to the fractional accuracy of $\sim 1\%$ compared to the $N$-body simulation results, over the range of scales up $k\lesssim 6\,h{\rm Mpc}^{-1}$ and in the range $0\le z\lesssim 1.5$, {\it if} the accurate response function is available. 
Encouragingly, even if we use the publicly available emulator of $P(k)$, which is calibrated for flat cosmologies (e.g. {\tt EuclidEmulator}
\cite{2021MNRAS.505.2840E}),
our method allows one to compute $P(k)$ for non-flat model, to within the accuracy of $\sim 2\%$. 

A key ingredient in our approach is that how the derivative operation with respect to $\Omega_K$ should be performed exactly in a multi-dimensional input parameter space under a constraint, $\Omega_{K} = 1 - (\Omega_\mathrm{m}+\Omega_\Lambda)$. In our case, the SU approach guides us to use $\delta_{\rm b}$ to fully specify the direction along which the derivative is taken. 
Then, we numerically find that this derivative coincides well with the derivative with respect to $h$ within flat cosmologies. 
This turns out to be practically useful to model non-flat cosmologies using only the knowledge within flat cosmologies. We can consider different ways to match flat-$\Lambda$CDM and \KCDM\, models, 
or more general models such as $w$CDM models with the equation-of-state parameter $w$ for dark energy.
It might be of interest to study more about the similarities and differences of the responses with respect to different combinations of parameters, as well as the time variable, at the nonlinear level beyond the applicable range of the one-to-one correspondence between $P(k)$ and $P^{\rm L}(k)$, which is valid within the EdS approximation and is explicitly used in methods such as \texttt{RESPRESSO}.
We will postpone to address a more comprehensive study in this direction as a future work.

An obvious application is to apply the method to actual data for constraining 
the curvature parameter $\Omega_K$. We will explore this direction in our future work. 
It is really interesting to explore a constraint of $\Omega_K$ from galaxy surveys, independently from the CMB constraint. 
As discussed in Ref.~\cite{2015PhRvD..92l3518T}, if we have precise BAO measurements at multiple redshifts (more than 3 redshifts), we 
can constrain $\Omega_K$ without employing any prior on the sound horizon (BAO scale) from CMB, because such multi-redshift BAO measurements can give sufficient information on the sound horizon scale and the cosmological distances that depend 
on $\Omega_{\rm m}$ and $\Omega_K$ ($\Omega_{\Lambda}$ is set by the identity $\Omega_K=1-[\Omega_{\rm m}+\Omega_{\Lambda}]$)
for non-flat $\Lambda$CDM models. However, note that the galaxy BAO geometrical constraints need to assume the existence of the standard ruler 
(i.e. BAO scale) over the multiple redshifts, as supported by the adiabatic initial condition, and the constraints would be degraded if employing further extended models such as time-varying dark energy models. In addition, we should try to explore the curvature information from the growth history of cosmic structures, in addition to the geometrical constraints. Once such high-precision constraint on $\Omega_K$ is obtained from galaxy surveys, 
we can address whether the {\it Planck} constraint and galaxy surveys, or more generally the late-time universe, are consistent with each other  within the adiabatic $\Lambda$CDM framework. Any deviation or inconsistency in these tests would be a smoking gun evidence of new physics beyond the standard 
$\Lambda$CDM model, and this will be definitely an interesting and important direction to explore with actual datasets. 

The response $T_{\delta_{\rm b}}$ is also a key quantity for calibrating the super sample covariance
(SSC), which is a dominant source of the non-Gaussian errors in the correlation functions of cosmic shear
\citep{sato09,2009MNRAS.395.2065T,2013MNRAS.429..344K,2013PhRvD..87l3504T}.
For future weak lensing surveys, it is important to obtain an accurate calibration of the non-Gaussian covariance, e.g. to have a proper assessment 
of the best-fit model compared with the statistical errors and not to have any significant bias in estimated parameters in the parameter inference \citep{2013MNRAS.432.1928T}. Estimating the SSC term for an arbitrary cosmological model is computationally expensive, because it requires to run a sufficient number of SU simulations (including the simulations in non-flat cosmologies) to have an accurate estimation of the SSC terms. 
For this, the approximation of $T_{\delta_{\rm b}}\approx T_h$ is also useful because we can use the public code of $P(k)$ to compute the SSC for a given cosmological model. However, note that, to model the total power of the SSC term, we further need to take into account the dilation effect \citep{2014PhRvD..89h3519L}, which is straightforward to compute from the numerical derivative of the nonlinear power spectra with respect to 
$k$. 

The SSC term is also significant or not negligible for galaxy-galaxy weak lensing or galaxy clustering, respectively \citep{2019MNRAS.482.4253T}.
The SSC terms for these correlation functions arise from the responses of the matter-galaxy or galaxy-galaxy power spectra, $P_{\rm gm}$ or $P_{\rm gg}$, 
to the super survey mode, $\delta_{{\rm b}}$. Since the galaxy-halo connection is modeled by the halo occupation distribution (HOD) model, it is interesting to study whether the normalized growth 
response of the matter-galaxy (matter-halo) or galaxy-galaxy (halo-halo) power spectra to 
$\delta_{\rm b}$ is approximated by the normalized response to $h$ similarly to the case for the matter power spectrum. This is our future work and will be presented elsewhere.

\acknowledgments
We would like to thank Kaz~Akitsu, Takahiko~Matsubara, Ravi~Sheth and 
Masato~Shirasaki for their useful and stimulating discussion. 
We thank the Yukawa Institute for Theoretical Physics at Kyoto University for their warm hospitality, where this work was partly done during the YITP-T-21-06 workshop on ``Galaxy shape statistics and cosmology''.
This work was supported in part by World Premier International Research Center Initiative (WPI Initiative), MEXT, Japan, JSPS KAKENHI Grant Numbers JP22H00130, JP20H05850, JP20H05855, 	JP20H05861, JP20H04723, JP19H00677, JP21H01081, JP22K03634, Basic Research Grant (Super AI) of Institute for AI and Beyond of the University of Tokyo, and Japan Science and Technology Agency (JST) AIP Acceleration Research Grant Number JP20317829.
Numerical computations were carried out on Cray XC50 at Center for Computational Astrophysics, National Astronomical Observatory of Japan.

\appendix

\section{Validation of the power spectrum responses with halo model}
\label{sec:halo_model}

In this section we 
study whether 
the approximate identity of $T_{\delta_{\rm b}}(k) \approx T_{h}(k)$ is reproduced by the halo model
\citep{Cooray02}.

\subsection{halo model approach}

The halo model gives a useful (semi-)analytical description of the nonlinear clustering statistics, and allows us to study the power spectrum responses 
\citep[see][for the similar study]{2013PhRvD..87l3504T,2014PhRvD..89h3519L,Chiang2014}.
In the halo model, the power spectrum is given by sum of the 1- and 2-halo terms as
\begin{align}
  P(k) = P^{\rm 1h}(k) + P^{\rm 2h}(k),   
\end{align}
where
\begin{align}
  P^{\rm 1h}(k) \equiv \int\!\mathrm{d}M~ n(M) \left(\frac{M}{\bar{\rho}_{\rm m}}\right)^2 \tu_M(k)^2
\end{align}
and
\begin{align}
  P^{\rm 2h}(k) \equiv \left[ I_1^1(k) \right]^2 P^{\rm L}(k).
\end{align}
with the function defined as
\begin{align}
  I_{\mu}^{\beta}(k_1,\cdots,k_{\mu}) \equiv \int\!\mathrm{d}M~ n(M) \left(\frac{M}{\bar{\rho}_{\rm m}}\right)^{\mu} b_{\beta}(M) \prod_{\i=1}^{\mu} \tu_M(k_i),
\end{align}
where $n(M)\mathrm{d}M$ is the number density of halos in the mass range $[M,M+\mathrm{d}M]$ (i.e. the halo mass function),  
$b_\beta(M)$ is the bias parameter for halos with mass $M$, 
defined in that $b_0=1$ and $b_1(M)$ is the linear bias parameter, 
and $\tilde{u}_M(k)$ is the Fourier transform of the mass density profile of halos with mass $M$. 
Note that the halo profile is normalized so as to satisfy 
$\tilde{u}_M(k)\rightarrow 1$ at $k\rightarrow 0$. 
With this normalization, $I^1_1(k)$ should be normalized at very small $k$ so as to satisfy the linear limit $I^1_1(k)\rightarrow 1$ at $k\rightarrow 0$ in that the 2-halo term reproduces the linear matter power spectrum, 
$P^{\rm 2h}(k)\simeq P^{\rm L}(k)$.
For the halo mass density profile, we
assume the Navarro-Frenk-White (NFW) halo profile \citep{NFW} in the following, where 
we estimate the halo concentration for halos in each mass bin from simulations.

We can formally express the power spectrum response with respect to a parameter $p$ ($p=\delta_{\rm b}$ or $h$) as  
\begin{align}
\frac{\partial P(k)}{\partial p}
=\frac{\partial P^{\rm 1h}(k)}{\partial p} + \frac{\partial P^{\rm 2h}(k)}{\partial p}
\end{align}
where the 1-halo term response is given as
\begin{align}
  \frac{\del P^{\rm 1h}(k)}{\del p}   = \int\!\mathrm{d}M~ n(M) \left(\frac{M}{\bar{\rho}_{\rm m}}\right)^2 \tu_M(k)^2 \nonumber \\ 
  \times \left[\frac{\del \ln n(M)}{\del p} +2\frac{\del \ln \tilde{u}_M(k)}{\del p}  \right].
\label{eq:1htermresp}
\end{align}
The 2-halo term response is given as
\begin{align}
  \frac{\del P^{2h}(k)}{\del p}   = 2I_1^1(k) \frac{\del I_1^1(k)}{\del p} P^{\rm L}(k) + \left[I_1^1(k)\right]^2 \frac{\del P^{\rm L}(k)}{\del p}. 
\label{eq:2htermresp}
\end{align}
Here the 2nd term, i.e. the linear power spectrum response $\partial P^{\rm L}/\partial p$, is equivalent to the response of the linear growth factor as discussed 
in Section~\ref{sec:spherical_collapse}. Hence it is straightforward to compute the 2nd term using the linear growth factor. 
Since the 2-halo term gives a dominant contribution to the total power in the linear regime, where $I^1_1\simeq 1$ as discussed above, we ignore the 1st term for the following results, for simplicity.

\subsection{Evaluation with $N$-body simulation}

In this section we use the $N$-body simulations in Table~\ref{tab:simulations} to calibrate
each term of the 1-halo term response (Eq.~\ref{eq:1htermresp}); 
more exactly, the responses of the halo mass function and the halo mass density profile. 

First we need to define halos from each output of $N$-body simulations. We follow the method in \citet{Nishimichi_2019}, so 
please see the paper for further details. As we emphasize around Table~\ref{tab:simulations}, all the $N$-body simulations employ the same box size 
($L\simeq 1.49\,{\rm Gpc}$), the same $N$-body particle number ($2048^3$) and the same $N$-body mass scale ($m_p=1.52\times 
10^{10}~M_\odot$). In this setting, the mean comoving mass density $\bar{\rho}_{\rm m0}$ is the same for all the simulations. 
To identify halos in each simulation output, we use the public software {\tt Rockstar} \citep{Behroozi:2013} that identifies halos and subhalos based on the clustering of $N$-body particles in phase space. For each halo/subhalo, we compute the spherical overdensity, $\Delta=200$, to define mass of each halo/subhalo in the comoving coordinates, $M=(4\pi/3)(R_{\rm 200m})^3\bar{\rho}_{\rm m0}\Delta$. 
Note that the halo mass definition is different from that used in the study of halo bias calibration using the SU simulations \citep[e.g.][]{2016PhRvD..93f3507L},
where the spherical overdensity is set to be $\Delta = 200/(1+\delta_{\rm b})$ in the 
SU simulation 
so that halos are identified using the same {\it physical}
overdensity as the corresponding global universe.
By using this halo definition, we can estimate only the ``growth'' response for the 1-halo term,
in the decomposition of ``growth'' and ``dilation'' responses \cite{2014PhRvD..89h3519L}.
Hence this response calibration is different from the method in \citep{2016PhRvD..93f3507L}.

After we identified halo candidates, we determine whether they are 
central or satellite halos. When the separation of two different halos (between their centers) is closer than $R_{\rm 200m}$ of the more massive one, we mark 
the less massive one as a satellite halo. In the following we use only central halos with mass containing more than 100~particles. With these definitions, 
each halo in all the simulations contains exactly the same number of member particles, which allows a cleaner calibration of the power spectrum responses 
in the halo model approach.

We first estimate the halo mass function from the halo catalog in each simulation realization. We use the following fitting function, which is a modified version of the earlier work in \citet{press74} \citep[also see][]{sheth02}, 
to fit the mass function estimated from the simulation: 
\begin{equation}
  n(M) \equiv \frac{\mathrm{d}n}{\mathrm{d}M} = f(\sigma_M)\frac{\bar{\rho}_{\rm m0}}{M} \frac{\mathrm{d}\ln \sigma_M^{-1}}{\mathrm{d}M},
\end{equation}
with 
\begin{equation}
  f(\sigma_M) = A\left[\left(\frac{\sigma_M}{b}\right)^a  + 1\right] \exp\left(-\frac{c}{\sigma_M^2}\right),
\end{equation}
where $A$, $a$, $b$ and $c$ are fitting parameters. The mass variance $\sigma_M^2$ is defined as
\begin{equation}
  \sigma_M(z)^2 \equiv 
  %\frac{1}{2\pi^2} 
  \int\!\frac{k^2\mathrm{d}k}{2\pi^2} 
  %k^2 
  P^{\rm L}(k,z) |\tW_R(k)|^2,
\end{equation}
where $\tW_R(k)$ is the Fourier transform of a top-hat filter of radius $R$ that is specified by an input halo mass 
via $R=(3M/4\pi\bar{\rho}_{\rm m0})^{1/3}$. 

\begin{figure}
	\includegraphics[width=0.45\textwidth]{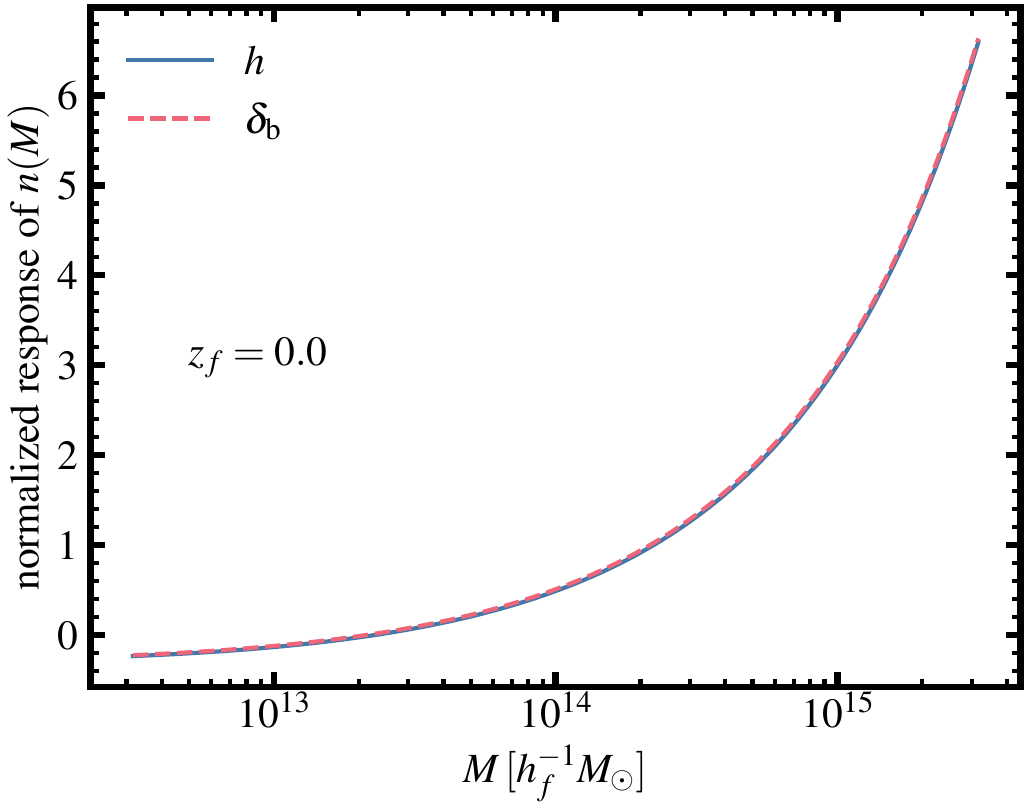}
	\caption{Normalized response of the halo mass function, $n(M)$, to $\delta_{\rm b}$ or
    $h$ at $z_{f}=0$, where the normalization was done in the same was as in Fig.~\ref{fig:Tdeltab,h}.
    These are computed from the halo catalogs in the $N$-body simulations for $\KCDM1$ and $\delta h$-$\Lambda$CDM models 
    in Table~\ref{tab:simulations} (see test for details). The two responses are in good agreement with each other. 
    }
	\label{fig:massfunctionResponse}
\end{figure}

For each of the simulations for $\KCDM1$ and $\delta h$-$\Lambda$CDM models in Table~\ref{tab:simulations},
we estimate the best-fit values of  $A$ and $a$ by fitting the above formula to the mass function measured from each simulation, assuming the Poisson noise in each halo mass bin. For the parameters $b$ and $c$, we fixed their values to those in \citet{Tinker08}. We use 10 realizations for each of the models with $\delta_{\rm b}=\pm 0.01$ at $z_f = 0$ ($\KCDM1$) 
and the plus or negative variations of $h$ from its fiducial value ($\delta h$-$\Lambda$CDM). We then estimate the responses of the halo mass function with respect to $\delta_{\rm b}$ or $h$ from the averaged mass function, using the two-side numerical derivatives: $\partial \ln n(M)/\partial \delta_{\rm b}$ or $\partial \ln n(M)/\partial h$,
which is the first term of the 1-halo term response (Eq.~\ref{eq:1htermresp}).
Note that this corresponds to the growth response of halo mass function due to the reason we described above. 
Fig.~\ref{fig:massfunctionResponse} shows the results for the mass function responses at $z_f=0$. Here we normalized the 
responses in the same way as those in $T_{\delta_{\rm b}}$ and $T_h$ using the responses of the linear growth factor 
(see around Eq.~\ref{eq:ps_estimator_tb}). The figure shows that the responses are in remarkably nice agreement with each other.  This agreement supports that the halo mass function is approximately 
given by a ``universal'' form, i.e. $f(\nu)$, where 
$\nu\equiv \delta_{\rm c}/\sigma_M(z)$ ($\delta_{\rm c}$ is a critical collapse threshold)
or $\nu\propto 1/\sigma_M(z)$, for different cosmological models. In this case, the halo mass function response is given by $\partial \ln n(M)/\partial p\propto \partial \ln f/\partial \nu\times \partial \nu/\partial p = - \partial \ln f/\partial \ln \nu\times \partial \ln D/\partial p$, %\tnrv{\bf (TN: Is this correct? I think that the intermediate expression is equal to $-\partial\ln f /\partial \ln \nu \times \partial \ln D / \partial p$, where the minus sign comes from the fact that $\sigma_M$ appears in the denominator in $\nu=\delta_c/\sigma_M$ (of course, you can omit this minus sign if you directly connect the most left-hand side to the final expression via ``$\propto$''). Please double check this.)}
since $\sigma_M(z)\propto D(z)$.
Note that we estimated the parameters $A$ and $a$ independently for different cosmological models, so the universality breaks down if 
the parameters $A$ and $a$ differ in the different models. 

\begin{figure}
	\includegraphics[width=0.45\textwidth]{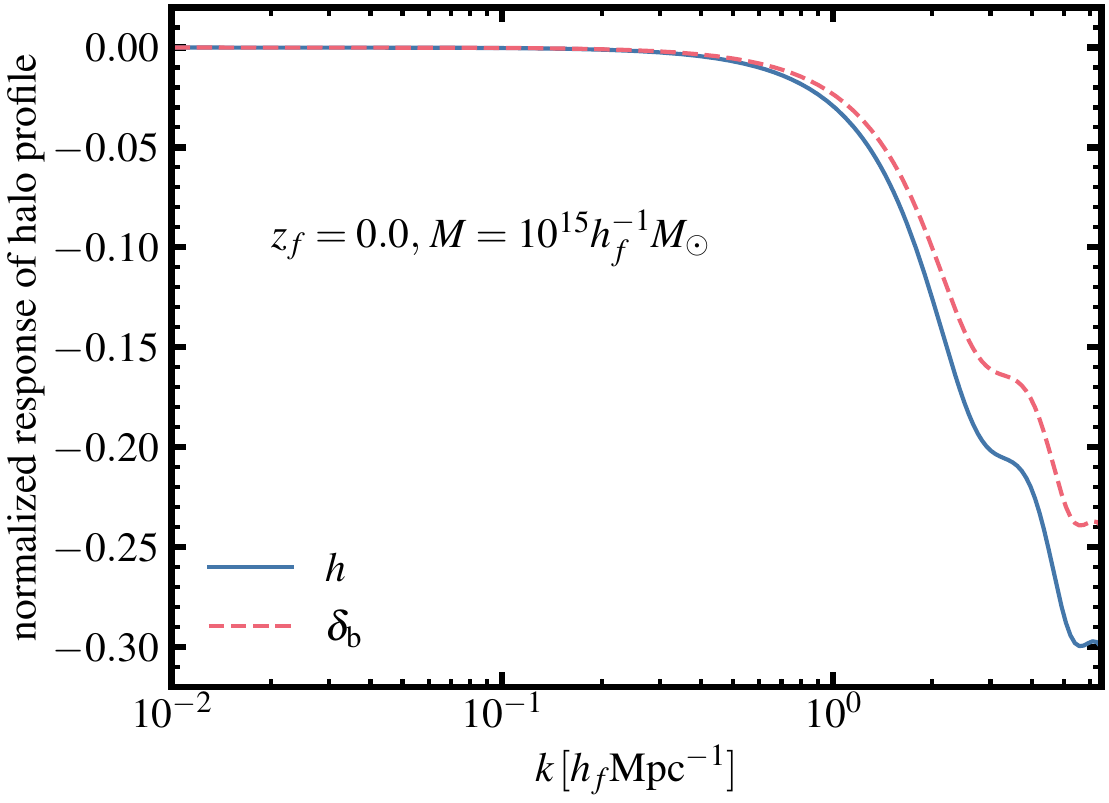}
	\caption{Normalized response of the Fourier transform of the halo mass density profile, 
    $\tilde{u}_M(k)$,  to $\delta_{\rm b}$ or $h$ for halos in the mass bin of $M=10^{15}~h_f^{-1}M_\odot$
    at $z_f=0$, as in the previous figure. To compute this, we first fit the averaged mass profile of halos 
    in the mass bin by an NFW profile to obtain the halo concentration in the simulations for each of the $\KCDM$1 
    and $\delta h$-$\Lambda$CDM model with varying $\delta_{\rm b}$ or $h$. Then we estimated the variations in the halo mass concentration between the varied cosmological models. The figure shows the results where we insert the 
    variations of the halo mass concentration into the NFW profile (see text for details).
    }
	\label{fig:profileResponse}
\end{figure}

Next we employ the following method to estimate the responses of the halo mass density profile, which is the 2nd term of Eq.~(\ref{eq:1htermresp}), in each simulation. We divide halos into each of 20 logarithmically-spaced mass bins 
in the range of $M=[10^{12.45},10^{15.45}]\,h_f^{-1}M_\odot$, and measure the ``averaged'' halo mass profile of halos in each bin. We fit each of the estimated mass profiles by an NFW profile to estimate the best-fit concentration parameter, assuming the Poisson errors according to the number of $N$-body particles contained in each of the radial bins. We then compare the best-fit NFW profiles to estimate the responses of the halo mass concentration with respect to the variations of $\delta_{\rm b}=\pm 0.01$ at $z_f = 0$ and $\delta h= \pm 0.02$, from the simulations for $\KCDM1$ and $\delta h$-$\Lambda$CDM models.
Fig.~\ref{fig:profileResponse} shows the responses of $\tilde{u}_M(k)$ with respect to $\delta_{\rm b}$ and $h$ for halos with $10^{15}h_f^{-1}M_\odot$ at $z_f=0$, where we employ the same normalization as in 
Fig.~\ref{fig:massfunctionResponse}. To estimate these responses, we plug in the variations of the halo concentration parameters into the Fourier transform of NFW profile. Note that the responses are by definition vanishing in small $k$ bins, where the normalized profile $u_M(k)=1$. The figure shows that the halo profile responses show a sizable difference 
at scales, where $u_M(k)<1$. The difference implies that the two responses do not exactly agree with each other 
at large $k$ in the nonlinear regime. This difference would be the origin of the slight discrepancy in $T_{\delta_{\rm b}}$ and $T_h$ at $k\gtrsim 1\,h_{f}{\rm Mpc}^{-1}$.

\begin{figure}
	\includegraphics[width=0.45\textwidth]{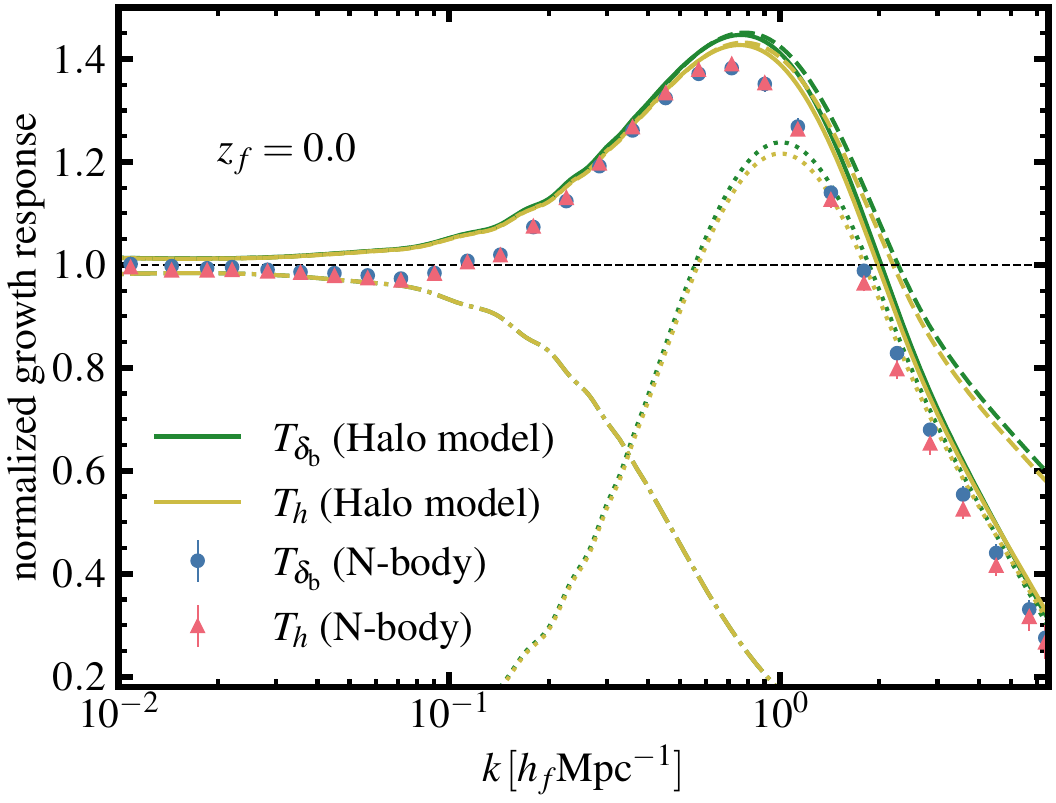}
	\caption{Solid lines show the halo model predictions for the normalized responses 
    of matter power spectrum to $\delta_{\rm b}$ and $h$, i.e. $T_{\delta_{\rm b}}$ and $T_h$,  at $z_f=0$. The dotted and dot-dashed
    lines show the 1- and 2-halo term predictions, respectively. The dashed lines denote the halo model predictions where we ignored the responses of the halo mass density profile or equivalently when we include only the halo mass function responses. For comparison, we show the simulation results for $T_{\delta_{\rm b}}$ and $T_h$ which are the same as those in the lower-right panel of Fig.~\ref{fig:Tdeltab,h}.
    }
	\label{fig:HalomodelResponse}
\end{figure}

Fig.~\ref{fig:HalomodelResponse} shows the normalized growth responses of matter power spectrum with 
respect to $\delta_{\rm b}$ and $h$, $T_{\delta_{\rm b}}(k)$ and $T_h(k)$, which are 
computed 
using the halo model: the sum of the 1-halo and 2-halo terms (Eqs.~\ref{eq:1htermresp} and \ref{eq:2htermresp}). 
To compute these results, we used the results of Figs.~\ref{fig:massfunctionResponse} and \ref{fig:profileResponse}. 
First, the figure clearly shows that the halo model predictions for $T_{\delta_{\rm b}}$ and $T_h$ 
agree well with each other. As expected, the scale-dependent responses arise from the 1-halo term, and 
the responses at $k\gtrsim 1\,h_{f}{\rm Mpc}^{-1}$ arises from the responses of the halo mass density profile. 
Thus these results give another confirmation of the approximate consistency of $T_{\delta_{\rm b}}$ and $T_h$ 
for $\Lambda$CDM model. However, the halo model cannot well reproduce the simulation results for the power spectrum response, especially at transition scales between the 1- and 2-halo terms, reflecting the limitation of the halo model. 

\bibliography{main.bbl}

\end{document}